\documentclass[a4paper,11pt]{article}
\pdfoutput=1 
\usepackage{jheppub2} 
\usepackage{bbm,bm,graphicx,mathtools,color,hyperref,slashed}
\usepackage{amssymb,amsmath}
\usepackage[T1]{fontenc}

\usepackage[all]{xy}

\graphicspath{{./Figures/}}

\newcommand{\diff}{\mathrm{d}}
\newcommand{\im}{\mathrm{i}}
\newcommand{\rmd}{\mathrm{d}}
\newcommand{\p}{\partial}
\newcommand{\Diff}{{\mathcal{D}}}

\newcommand{\rme}{\mathrm{e}}

\preprint{YITP-25-78}

\title{2d Cardy-Rabinovici model with the modified Villain lattice: Exact dualities and symmetries}

\author{Nagare Katayama, }

\emailAdd{nagare.katayama@yukawa.kyoto-u.ac.jp}

\author{Yuya Tanizaki}

\affiliation{Yukawa Institute for Theoretical Physics, Kyoto University, Kyoto 606-8502, Japan}

\emailAdd{yuya.tanizaki@yukawa.kyoto-u.ac.jp}

\abstract{
The Cardy-Rabinovici model is a toy model of the lattice $U(1)$ gauge theories to study various oblique confinement states associated with the nonzero $\theta$ angles. 
We reformulate the $2$d version of this model using the modified Villain lattice formalism, and we establish the exact $\theta$ periodicity for the Witten effect and the strong-weak duality at the finite lattice spacings. 
We then study the phase structure of this model based on the duality, symmetry and anomaly, and the perturbative renormalization group. 
}

\begin{document}
\maketitle
\section{Introduction}
\label{sec:introduction}

Quantum chromodynamics (QCD) is the fundamental theory of strong interactions, whose elementary degrees of freedom are quarks and gluons. 
We can only observe hadrons in the vacuum of our universe, and quarks and gluons are confined inside color-singlet hadrons. 
Magnetic excitations, such as monopoles and center vortices, play the pivotal role in describing color confinement in $4$d gauge theories~\cite{Nambu:1974zg, Mandelstam:1974pi, Polyakov:1975rs, tHooft:1977nqb, tHooft:1979rtg, Cornwall:1979hz, Nielsen:1979xu, Ambjorn:1980ms}: For example, if color magnetic monopoles condense in the vacuum, then color electric charges should be confined as their electric fluxes are collimated and forming the flux tube via the dual Meissner effect. 
The importance of solitonic objects makes the physics of nonperturbative gauge theories quite intriguing, while developing their theoretical descriptions becomes challenging.

$4$d gauge theories have the $\theta$ angle associated with the instanton number, which provides an efficient probe for the dynamics of magnetic excitations. 
As a general remark, the set of possible electric and magnetic charges in $4$d gauge theories is highly constrained by the mutual locality condition to make Dirac strings invisible~\cite{Dirac:1931kp, Schwinger:1966nj, Zwanziger:1968rs}. 
Still, it turns out that there exists one free parameter for the electric charge of the magnetic monopole, and the $\theta$ parameter controls it through the Witten effect: The magnetic monopole of magnetic charge $m$ acquires the fractional electric charge $\frac{\theta}{2\pi}m$ at nonzero values of $\theta$ and becomes a dyon~\cite{Witten:1979ey}. 
The $\theta$ angle is a $2\pi$ periodic parameter for the path-integral weight, but the Witten effect clarifies the nontrivial meaning of its periodicity. The $2\pi$ periodicity of $\theta$ implies the unitary equivalence for quantum systems at $\theta$ and $\theta+2\pi$, and the unitary transformation between them reshuffles the electric charges of the dyon spectrum.

In confinement phases, the $\theta$ angle has a drastic effect on the vacuum properties as it directly affects condensing magnetic particles, which opens the possibility for fruitful varieties of confinement phases called oblique confinement~\cite{tHooft:1981bkw}. 
Cardy and Rabinovici proposed a toy model of oblique confinement phases based on the lattice $U(1)$ gauge theory~\cite{Cardy:1981qy, Cardy:1981fd}, which we call the Cardy-Rabinovici (CR) model in this paper. 
The CR model introduces both electric and magnetic matters by summing up ensembles of charge-$N$ Wilson and charge-$1$ 't~Hooft loops with the $\theta$-dependent lattice Coulomb interaction, and Cardy and Rabinovici conjectured its phase diagram by giving an ansatz for the one-particle free energy of the condensing particle.
The proposed phase diagram of this model has been recently revisited from the viewpoint of the generalized global symmetry, giving the reinterpretation of oblique confinement phases as topological states of quantum matter~\cite {Ye:2016ase, Honda:2020txe, Hayashi:2022fkw, Moy:2022ztf, Pace:2022cnh, Moy:2024cgm}. 

Thus, there is no doubt that the CR model contains the essence of many nontrivial aspects of confinement. 
However, the original proposal of the CR model suffers from several serious problems: 
\begin{enumerate}
    \item The lattice $\theta$ angle does not respect the $2\pi$ periodicity of the continuum formulation. 
    \item The locality of the model is not manifest as the matter part does not take the form of local fields but is realized as ensembles of loop operators. 
\end{enumerate}
Until recently, people did not have a way to circumvent these problems for the lattice regularized $U(1)$ gauge theories. 
The seminal works by Gattringer, Sulejmanpasic, and their collaborators~\cite{Gattringer:2018dlw, Sulejmanpasic:2019ytl, Sulejmanpasic:2020lyq, Anosova:2022cjm, Anosova:2022yqx, Jacobson:2023cmr} develop the modified Villain formulation, which can resolve the above issues at a time. 
The Villain formulation~\cite{Villain:1974ir} of the lattice $U(1)$ gauge theories uses the $\mathbb{R}$-valued link variables and also the $\mathbb{Z}$-valued plaquette variables to describe the photon fluctuations and the Dirac-quantized magnetic fluxes separately, which has a sharp contrast with the Wilson formulation using the $U(1)$-valued link variables that represent both. 
The important discovery of Ref.~\cite{Sulejmanpasic:2019ytl} is that one should also introduce the $\mathbb{R}$-valued dual link variable as the Lagrange multiplier forcing the monopole-less condition by its equation of motion to control monopole dynamics. 
The modified Villain formulation is applied to the lattice construction of fracton models~\cite{Gorantla:2021svj} and of non-invertible symmetry generators~\cite{Choi:2021kmx}, and of $2$d chiral gauge theory through Bosonization~\cite{Berkowitz:2023pnz}, and its idea is further extended to give the lattice definition of the non-Abelian topological charges~\cite{Chen:2024ddr, Zhang:2024cjb}.  
Then, it is an interesting problem to reformulate the CR model based on the modified Villain formulation as it would enable us to see explicitly how the topology affects the quantum dynamics while controlling the ultraviolet divergences. 
In particular, we would like to note that Ref.~\cite{Anosova:2022cjm} constructed a self-dual $4$d $U(1)$ lattice gauge theory with the $\theta$ term using the modified Villain formulation, which gives the foundation to reformulate the CR model.

In this paper, let us consider the $2$d version of the CR model and develop its modified lattice formulation to study its dynamics in detail. 
The $2$d CR model is obtained by the dimensional reduction of the $4$d CR model by putting it on $\mathbb{R}^2\times T^2$~\cite{Cardy:1981qy, Cardy:1981fd}, and the $U(1)$ gauge fields for the compactified $T^2$ direction become the $2$d compact bosons. 
The $2$d two-component compact boson has the $\theta$ angle, which comes out of the $\theta$ angle of the $4$d Maxwell theory, and the monopole singularities in $4$d become the vortex singularities in the $2$d setup. 
We can translate all the discussions about the $4$d CR model to the ones for the $2$d CR model, including the Witten effect, the conjectured phase diagram, and also the problems appearing in the lattice regularization. 
Therefore, the modified Villain formulation should be suitable for retaining the continuum knowledge as much as possible. 

In Sec.~\ref{sec:2d compact boson theory with theta term}, we give the modified Villain formulation of the $2$d free compact boson with the $\theta$ term after a brief review of the continuum description. 
We will see the exact lattice realizations of the Witten effect and the strong-weak duality. 
In Sec.~\ref{sec:{2d Cardy-Rabinovici model with modified Villain lattice formulation}}, we construct the $2$d Cardy-Rabinovici model that exactly enjoys the (rescaled) $\theta$ periodicity and the strong-weak duality associated with the $\mathbb{Z}_N\times \mathbb{Z}_N$ gauging. 
We find the 't~Hooft anomalies at the finite lattice spacings, and we analyze the phase structure based on the duality and symmetry combined with the perturbative renormalization group. 
We summarize the result and discuss some future directions in Sec.~\ref{sec:summary}.

\section{\texorpdfstring{$2$d}{2d} free compact boson theory with \texorpdfstring{$\theta$}{theta} term: continuum and lattice}
\label{sec:2d compact boson theory with theta term}

In this section,  we consider the two-component $2$d free compact boson theory with $\theta$ term. 
We first give its brief review in its continuum formulation and see how the Witten effect is realized in this system. 
Then, we give the modified Villain formulation of this two-component free compact boson and realize the Witten effect exactly at finite lattice spacings.

\subsection{Continuum description of the \texorpdfstring{$\theta$}{theta} angle and Witten effect}
\label{subsec:{Continuum description of the theta angle and Witten effect}}

Let us consider a $2$d free scalar theory with two compact bosons, $\phi_a:M_2\to \mathbb{R}/2\pi \mathbb{Z}$ $(a=1,2)$, where $M_2$ is a $2$d closed Euclidean spacetime. 
When we think of $\phi_a(x)$ as a locally $\mathbb{R}$-valued field, each $\phi_{a}(x)$ enjoys the identification
\begin{align}
    \phi_{a}(x)
    \sim
    \phi_{a}(x) + 2\pi,
\end{align}
that is, all the physical observables have to be invariant under this gauge redundancy. 
Thus, $\phi_a$ itself is not physical, and the physical ones have to be exponentiated as 
\begin{equation}
    \exp(\im n \phi_a(x)), 
\end{equation}
with some $n\in \mathbb{Z}$, which is called the vertex operator in the string theory context.  There also exists the dual vertex operator, or the vortex operator, 
\begin{equation}
    \exp(\im m \tilde{\phi}_a(y)), 
\end{equation}
which introduces the $2\pi m$ winding singularity for $\phi_a(x)$ at $x=y$: When inserting this operator, we require that $\phi_a(x)=m \arg((x_1-y_1)+\im (x_2-y_2))+(\text{regular part})$ in the vicinity of $y$. 
We note that the branch cut of ``$\arg$'' cannot be observed by physical observables as $n,m\in \mathbb{Z}$. 
The notation of $\tilde{\phi}_a$ might be mysterious at this stage but will be clarified later when we look at the modified Villain lattice formulation of this system. 

As the Euclidean free-theory action, let us consider\footnote{
Let us make its connection with the $4$d Maxwell theory mentioned in Introduction. 
We write $A$ for the $4$d $U(1)$ gauge field, and the $4$d Maxwell action with the $\theta$ term is $S=\int_{M_4}\left[\frac{1}{2e^2}|\diff A|^2+\im \frac{\theta}{8\pi^2}(\diff A)^2\right]$. 
We take the $T^2$ compactification, $M_4=M_2\times T^2\ni (x,(y_1,y_2))$, and decompose the $4$d gauge field as $A=A_{2\rmd}+\phi_1 \frac{\diff y_1}{L_A}+\phi_2 \frac{\diff y_2}{L_B}$, where $A_{2\rmd}$ is the $2$d $U(1)$ gauge fields and $\phi_{1,2}$ are holonomy fields along the compactified $T^2$ direction, $y_1\sim y_1+L_A$ and $y_2\sim y_2+L_B$. 
For simplicity, let us neglect $A_{2\rmd}$ in the following, and then we obtain the $2$d compact boson action~\eqref{eq:2d_CPT+theta_action} with $R_1^2=\frac{2\pi}{e^2}\frac{L_B}{L_A}$ and $R_2^2=\frac{2\pi}{e^2}\frac{L_A}{L_B}$. } 
\begin{align}
    S_{\theta}[\phi_{1} , \phi_{2}]
    =
    \int_{M_2} 
    \sum_{a=1,2} \frac{R_a^2}{4\pi} |\diff\phi_{a}|^2
    + \im \frac{\theta}{4\pi^{2}}\int_{M_2}\diff\phi_{1}\wedge \diff\phi_{2},
    \label{eq:2d_CPT+theta_action}
\end{align}
where $|\diff \phi_a|^2=\diff \phi_a\wedge \star \diff \phi_a=(\partial_\mu \phi_a)^2 \diff^2 x$, $R_a$ is the compact-boson radius, and the last term describes the topological $\theta$ term. The $\theta$ parameter couples to the $\mathbb{Z}$-quantized topological charge, 
\begin{align}
    Q_{\text{top}}
    =
    \frac{1}{4\pi^{2}}\int_{M_2}\diff\phi_{1}\wedge \diff\phi_{2},
    \label{top-term_continuum}
\end{align}
and the path-integral weight $\exp(-S_{\theta}[\phi_{1,2}])$ has the periodicity $\theta\sim \theta+2\pi$. 
Thus, the quantum system defined at $\theta$ and $\theta+2\pi$ should be unitary equivalent, but we note that they do not need to be identical on the nose, which is deeply related to the Witten effect~\cite{Witten:1979ey}. 

To see the Witten effect of this system, let us compute the two-point function of the vertex/vortex operators on $\mathbb{R}^2$ by taking the infinite-volume limit. The detailed computation is given in Appendix~\ref{appendix_A}, and the result is 
\begin{align}
    &\big\langle \rme^{\im (N_{1}\phi_{1}+ N_{2}\phi_{2}+ \tilde{N}_{1}\tilde{\phi}_{1}+ \tilde{N}_{2}\tilde{\phi}_{2})(z,\bar{z})}\rme^{-\im (N_{1}\phi_{1} + N_{2}\phi_{2} + \tilde{N}_{1}\tilde{\phi}_{1} + \tilde{N}_{2}\tilde{\phi}_{2})(0,0)}
    \big\rangle_{\theta} \notag\\
    &=
    \bigg(
    \frac{1}{|z|^{2}}
    \bigg)^
    {
    \frac{1}{2 R_1^2}\big(N_{1}+\frac{\tilde{N}_{2}\theta}{2\pi}\big)^{2}+\frac{R_1^2}{2}\tilde{N}_{1}^{2}
    + \frac{1}{2R_2^2}\big(N_{2}-\frac{\tilde{N}_{1}\theta}{2\pi}\big)^{2} + \frac{R_2^2}{2}\tilde{N}_{2}^{2}
    }
    \times
    \bigg(
    \frac{\bar{z}}{z}
    \bigg)^
    {N_{1}\tilde{N}_{1} + N_{2}\tilde{N}_{2}},
\end{align}
where $z=x_1+\im x_2$ is the complex coordinate. The first factor describes the contribution of the $2$d massless Coulomb potential, and the second term gives the spin contribution. The Coulomb-potential part gives the scaling dimension of the operator, 
\begin{align}
    &\quad \Delta[\rme^{\im (N_{1}\phi_{1}+ N_{2}\phi_{2}+ \tilde{N}_{1}\tilde{\phi}_{1}+ \tilde{N}_{2}\tilde{\phi}_{2})}] \notag\\
    &= \frac{1}{2R_1^2}\left(N_{1}+\frac{\theta}{2\pi}\tilde{N}_{2}\right)^{2} + \frac{R_1^2}{2}\tilde{N}_{1}^{2} 
    + \frac{1}{2R_2^2}\left(N_{2}-\frac{\theta}{2\pi}\tilde{N}_{1}\right)^{2} + \frac{R_2^2}{2}\tilde{N}_{2}^{2}. 
    \label{scaling-dim}
\end{align}
The coefficient of $1/R_a^2$ corresponds to the squared of the effective electric charge, and the coefficient of $R_a^2$ corresponds to the squared of the magnetic charge in the context of the $4$d Maxwell theory. 
The effective electric charge acquires the fractional contribution from the magnetic charge by $\frac{\theta}{2\pi}$, which is nothing but the Witten effect~\cite{Witten:1979ey}. 
The result  also shows that the local operator spectrum is reshuffled under the $2\pi$ shift of the $\theta$ angle as follows: 
\begin{equation}
    \rme^{\im \bigl(N_{1}\phi_{1}+ N_{2}\phi_{2}+ \tilde{N}_{1}\tilde{\phi}_{1}+ \tilde{N}_{2}\tilde{\phi}_{2}\bigr)} \text{ at $\theta+2\pi$} \, 
    \longleftrightarrow \,
    \rme^{\im \bigl((N_{1}+\tilde{N}_2)\phi_{1}+ (N_{2}-\tilde{N}_1)\phi_{2}+ \tilde{N}_{1}\tilde{\phi}_{1}+ \tilde{N}_{2}\tilde{\phi}_{2}\bigr)} \text{ at $\theta$},
    \label{correlation_theta-theta+2pi_continuum}
\end{equation}
which is caused by the nontrivial action of the unitary equivalence between $\theta+2\pi$ and $\theta$. 

Let us discuss the symmetry and its 't~Hooft anomaly of this system. There exists two electric $U(1)$ symmetries and also two magnetic $U(1)$ symmetries: 
\begin{align}
    \text{electric $U(1)$}:&\, \phi_a\mapsto \phi_a+\alpha_a \text{ with Noether current } j^{(a)}_{\mathrm{ele}} = \star \frac{R_a^2}{2\pi} \diff\phi_{a}(x), 
    \label{eq:electricU(1)}\\
    \text{magnetic $U(1)$}:&\, \tilde{\phi}_a\mapsto \tilde{\phi}_a+\tilde{\alpha}_a \text{ with Noether current } j^{(a)}_{\mathrm{mag}} = \frac{1}{2\pi} \diff\phi_{a}(x). 
    \label{eq:magneticU(1)}
\end{align}
With the background gauge fields $A_a, \tilde{A}_a$ for these symmetries, the gauged action becomes 
\begin{align}
    S_{\theta,\, \mathrm{gauged}}[\phi_a, A_a, \tilde{A}_a]
    &=
    \int_{M_2} 
    \sum_{a=1,2} \frac{R_{a}^2}{4\pi}|\diff\phi_{a} + A_a|^2
    + \im \frac{\theta}{4\pi^{2}}\int_{M_2}(\diff\phi_{1}+A_1)\wedge (\diff\phi_{2}+A_2) \notag\\
    &\quad +\im \sum_{a=1,2}\int_{M_2}\tilde{A}_a\wedge \frac{1}{2\pi}(\diff \phi_a+A_a). 
\end{align}
This action violates the background gauge invariance for the magnetic $U(1)$ symmetries, $\tilde{A}_a\mapsto \tilde{A}_a+\diff \lambda_a$, which needs to be cancelled by the $3$d anomaly-inflow action, 
\begin{equation}
    S_{\text{3d inflow}}[A_a,\tilde{A}_a]=\frac{\im}{2\pi}\int_{M_3}\sum_{a=1,2} \tilde{A}_a\wedge \diff A_a,
\end{equation}
with $\partial M_3=M_2$. This represents the mixed 't~Hooft anomaly between the electric and magnetic $U(1)$ symmetry. 

Let us note that the gauged action also violates the $2\pi$ periodicity of the $\theta$ parameter: 
\begin{equation}
    S_{\theta+2\pi, \, \text{gauged}}-S_{\theta, \, \text{gauged}}= \frac{\im}{2\pi}\int_{M_2} (\diff \phi_1 + A_1)\wedge (\diff \phi_2 + A_2). 
\end{equation}
This violation is more severe than the one of the 't~Hooft anomaly as there are mixed terms between the dynamical and background fields, such as $\frac{\im}{2\pi}A_1\wedge \diff \phi_2$. 
To eliminate the mixed terms, we may combine the $2\pi$ shift of $\theta$ with 
\begin{equation}
    \tilde{A}_1\mapsto \tilde{A}_1 + A_2,\quad \tilde{A}_2 \mapsto \tilde{A}_2 - A_1, 
\end{equation}
and then we obtain
\begin{align}
    &\qquad S_{\theta+2\pi, \, \text{gauged}}[\phi_a, A_a, \tilde{A}_a + \varepsilon_{ab} A_b]-S_{\theta, \, \text{gauged}}[\phi_a, A_a, \tilde{A}_a]\notag\\
    &= \frac{\im}{2\pi}\int_{M_2} (\diff \phi_1 + A_1)\wedge (\diff \phi_2 + A_2) +\frac{\im}{2\pi}\int_{M_2} [A_2\wedge (\diff \phi_1+A_1) - A_1\wedge (\diff \phi_2+ A_2)]\notag\\
    &=-\frac{\im}{2\pi}\int_{M_2} A_1\wedge A_2 \quad (\bmod \, 2\pi\im\,). 
\end{align}
This depends only on the background gauge fields, which is again canceled by the above $3$d anomaly-inflow action~\cite{Abe:2023uan}. 
We may interpret that the additional transformation, $\tilde{A}_a\mapsto \tilde{A}_a+\varepsilon_{ab}A_b$, represents the reshuffling~\eqref{correlation_theta-theta+2pi_continuum} associated with the Witten effect.

\subsection{Lattice description of the \texorpdfstring{$\theta$}{theta} angle and the Witten effect}
\label{subsec:{Lattice description of the theta angle and the Witten effect}}

In this section, we describe the lattice realization of the free compact boson using the modified Villain lattice formulation developed in Refs.~\cite{Gattringer:2018dlw, Sulejmanpasic:2019ytl, Sulejmanpasic:2020lyq, Anosova:2022cjm, Anosova:2022yqx, Jacobson:2023cmr}. 
In particular, we will demonstrate that the modified Villain formulation of the 2d compact boson theory realizes the Witten effect exactly at the finite lattice spacings.

\subsubsection{Construction of the modified Villain lattice action}
\label{subsubsec:{Construction of the modified Villain lattice action}}

Let us approximate $2$d torus spacetime $T^{2}$ as the 2d square lattice $\Gamma = \big(\mathbb{Z}/L \mathbb{Z}\big)^{2}$ of size $L$ with the periodic boundary condition. 
We write the lattice sites as $x,y,\ldots \in \Gamma$, the directions as $\mu,\nu,\ldots \in \{1,2\}$, and the unit vector along the $\mu$ direction as $\hat{\mu}$. 
We also use the dual lattice sites defined by $\tilde{x}=x+\frac{1}{2}(\hat{1}+\hat{2})$. 

\begin{figure}
    \centering
    \includegraphics[width=0.35\linewidth]{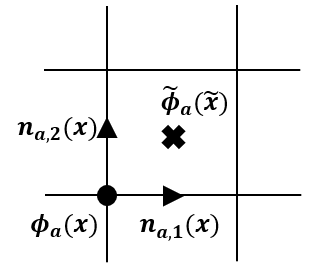}
    \caption{
    Illustration for the dynamical variables in the modified Villain lattice formulation. The scalar $\phi_a(x)\in (-\pi,\pi]\subset \mathbb{R}$ lives on the site $x$, the discrete gauge field $n_{a,\mu}(x)\in \mathbb{Z}$ lives on the link $(x,\mu)$, and the dual scalar $\tilde{\phi}_a(\tilde{x})\in (-\pi,\pi]\subset \mathbb{R}$ lives on the dual site $\tilde{x}=x+\frac{1}{2}(\hat{1}+\hat{2})$. }
    \label{fig:2d CPT-MVlattice-action}
\end{figure}

In the standard (Wilson-type) lattice formulation, we assign the $U(1)$-valued field, $\exp(\im \phi_a(x))\in U(1)$, on each site $x\in \Gamma$ as the lattice discretization of the compact boson. 
In the Villain formulation~\cite{Villain:1974ir}, we instead introduce the $\mathbb{R}$-valued site variables and the $\mathbb{Z}$-valued link variables, 
\begin{equation}
    \phi_a(x)\in \mathbb{R}, \quad n_{a,\mu}(x)\in \mathbb{Z},  
\end{equation}
and we impose the $\mathbb{Z}$-valued gauge invariance to represent $U(1)\simeq \mathbb{R}/2\pi \mathbb{Z}$, which will be described in more detail soon later. 
In the modified Villain formulation~\cite{Gattringer:2018dlw, Sulejmanpasic:2019ytl, Sulejmanpasic:2020lyq, Anosova:2022cjm, Anosova:2022yqx, Jacobson:2023cmr}, we also introduce the $\mathbb{R}$-valued Lagrange-multiplier field on the dual site (see Fig.~\ref{fig:2d CPT-MVlattice-action}), 
\begin{equation}
    \tilde{\phi}_a(\tilde{x})\in \mathbb{R},
\end{equation}
to impose the flatness condition on each integer gauge field $n_{a,\mu}$. This allows us to control the dynamics of vortices (or monopoles in the context of the $4$d Maxwell theory), which turns out play the pivotal role to realize the Witten effect with the $\theta$ angle on the lattice. 

\begin{figure}
    \centering
    \includegraphics[width=0.45\linewidth]{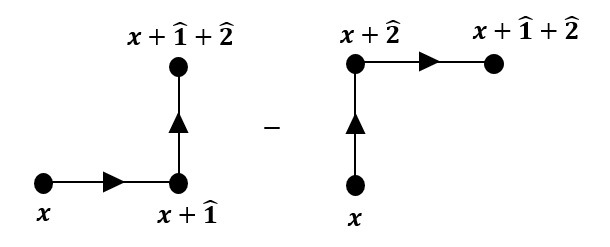}
    \caption{Illustration for the structure of the lattice definition of $Q_{\mathrm{top}}$. When discretizing the continuum expression $\int \frac{1}{4\pi^2}\mathrm{d}\phi_1\wedge \mathrm{d}\phi_2$, the location for $\diff \phi_2$ starts from the endpoint of $\diff \phi_1$. }
    \label{fig:2d CPT-MVlattice-topterm}
\end{figure}

The modified Villain lattice action for the free theory \eqref{eq:2d_CPT+theta_action} is given by 
\begin{align}
    S_{\mathrm{free},\, \theta}
    &=
    \sum_{a=1,2}\sum_{x,\mu}  \frac{R_a^2}{4\pi}[\partial_{\mu}\phi_{a}(x) + 2\pi n_{a,\mu}(x)]^{2}
    +\im \sum_{a=1,2}\sum_{x,\mu,\nu}\varepsilon_{\mu\nu}\tilde{\phi}_{a}(\tilde{x}) \partial_{\mu} n_{a,\nu}(x)
    \notag\\
    &\quad + \im \theta Q_{\text{top}}[\phi_a,n_{a,\mu}],
    \label{eq:latticeVillainAction_free}
\end{align}
where $\partial_{\mu}$ represents the forward difference in $\mu$-direction, $\partial_{\mu}\phi(n) = \phi(n+\hat{\mu})-\phi(n)$, and we define the lattice topological charge $Q_{\text{top}}$ by 
\begin{align}
    Q_{\text{top}}[\phi_a,n_{a,\mu}]
    =
    \frac{1}{4\pi^{2}}
    \sum_{x,\mu,\nu} \varepsilon_{\mu\nu}[
    \partial_{\mu}\phi_{1}(x) + 2\pi n_{1,\mu}(x)
    ]
    [
    \partial_{\nu}\phi_{2}(x+\hat{\mu}) + 2\pi n_{2,\nu}(x+\hat{\mu})
    ].
    \label{eq:lattice_Qtop}
\end{align} 
We illustrate its structure in Fig.~\ref{fig:2d CPT-MVlattice-topterm}. 
In the following, let us identify and confirm the gauge invariance of this system, and then we discuss the integer quantization of the lattice topological charge.

The lattice Boltzmann weight $\exp(-S_{\theta})$ is invariant under the following local $\mathbb{Z}$-valued transformations, 
\begin{align}
    &\left\{
    \begin{array}{l}
      \phi_{a}(x) \mapsto  \phi_{a}(x) - 2\pi k_{a}(x), \\ 
       n_{a,\mu}(x) \mapsto n_{a,\mu}(x) + \partial_{\mu}k_{a}(x),  
    \end{array}\right. 
    \label{eq:electric_gauge_trans}\\
    &\quad \tilde{\phi}_{a}(\tilde{x})
    \mapsto \tilde{\phi}_{a}(\tilde{x}) - 2\pi \tilde{k}_{a}(\tilde{x}).
    \label{eq:magnetic_gauge_trans}
\end{align}
where $k_{a}(x), \tilde{k}_a(x)\in \mathbb{Z}$. The physical observables are given by the vertex/vortex operators, 
\begin{equation}
    \exp(\im \phi_a(x)), \quad \exp(\im \tilde{\phi}_a(\tilde{x})), 
\end{equation}
instead of $\phi_a$, $\tilde{\phi}_a$. The electric and magnetic $U(1)$ symmetries \eqref{eq:electricU(1)}, \eqref{eq:magneticU(1)} act on these operators. 
One can fix the gauge by imposing the condition $-\pi<\phi_a(x),\tilde{\phi}_a(\tilde{x})\le \pi$, and the partition function is given by the path integral, 
\begin{align}
    Z_{\mathrm{free}}
    =
    \int \Diff\phi_{a}\int \Diff\tilde{\phi}_{a} \prod_{x,\mu} \sum_{n_{a,\mu}(x)\in \mathbb{Z}}
    \exp(-S_{\mathrm{free},\, \theta}[\phi_a,\tilde{\phi}_a,n_{a,\mu}]),
    \label{eq:latticeZ_free}
\end{align}
with the path-integral measure
$\int \Diff\phi_{a}
    =\prod_{x}\int_{-\pi}^{\pi}\frac{\diff\phi_{a}(x)}{2\pi},\,
\int \Diff\tilde{\phi}_{a}
    = \prod_{\tilde{x}}\int_{-\pi}^{\pi}\frac{\diff\tilde{\phi}_{a}(\tilde{x})}{2\pi}$. 

Lastly, let us comment on the Euclidean $\frac{\pi}{2}$ rotation of the square lattice.  
The $\theta$ term explicitly violates the naive $\frac{\pi}{2}$ rotation, but one can check there exists a nontrivial $\frac{\pi}{2}$ rotational symmetry by rotating the $\phi_1$ field around the site $x_*$ and the $\phi_2$ field around its dual site $\tilde{x}_*$. 
See Appendix~\ref{appendix_B} for more detailed discussion.
This property is important to constrain the local counterterms when considering the renormalization group. 

\subsubsection{Exact realization of the Witten effect on the lattice}
\label{subsubsec:{Exact realization of the Witten effect on the lattice}}

Let us confirm the integer quantization of the lattice topological charge $Q_{\text{top}}$. It turns out to be essential that the site for $\phi_2$ in \eqref{eq:lattice_Qtop} is shifted by $\hat{\mu}$ for the topologicalness of $Q_{\mathrm{top}}$.\footnote{If we use the lattice differential form and the cup product on the square lattice~\cite{Tata:2020qca, Chen:2021ppt}, \eqref{eq:lattice_Qtop} becomes $Q_{\mathrm{top}}=\frac{1}{(2\pi)^2}\int (\diff \phi_1+2\pi n_1)\cup (\diff \phi_2+2\pi n_2)$ thanks to the shift of the location of $\phi_2$ by $\hat{\mu}$. See also \cite{Dimakis:1992pk, Aschieri:1992wg}. } 

Let us expand the expression~\eqref{eq:lattice_Qtop} as 
\begin{align}
    Q_{\text{top}}
    &=
     \sum_{x,\mu,\nu}\varepsilon_{\mu\nu}
    \left[
    \frac{1}{4\pi^{2}}\partial_{\mu}\phi_{1}(x)\partial_{\nu}\phi_{2}(x+\hat{\mu})
    +\frac{1}{2\pi} n_{1,\mu}(x)\partial_{\nu}\phi_{2}(x+\hat{\mu}) \right.\notag\\
    &\quad\quad\quad \left. + \frac{1}{2\pi} \partial_{\mu}\phi_{1}(x)n_{2,\nu}(x+\hat{\mu})
    + n_{1,\mu}(x)n_{2,\nu}(x+\hat{\mu})
    \right].
    \label{eq:Qtop_expansion}
\end{align}
The last term is manifestly $\mathbb{Z}$-valued. 
To see other terms can be eliminated, we should note that the ``integration-by-part'' for the forward lattice derivative takes the following form, 
\begin{equation}
    \sum_x (\partial_\mu f(x)) g(x)=-\sum_x f(x)\partial_\mu g(x-\hat{\mu}).
\end{equation} 
Therefore, the first term vanishes,
\begin{equation}
    \sum_{x,\mu,\nu}\varepsilon_{\mu\nu}\partial_{\mu}\phi_{1}(x)\partial_{\nu}\phi_{2}(x+\hat{\mu})=-\sum_{x,\mu,\nu}\varepsilon_{\mu\nu}\phi_{1}(x)\partial_{\mu}\partial_{\nu}\phi_{2}(x)=0, 
\end{equation} 
since $\varepsilon_{\mu\nu}$ is antisymmetric in $\mu\nu$ while the other part is symmetric after the integration-by-parts. 
The second term can be rewritten as
\begin{align}
    \frac{\theta}{2\pi}\sum_{x,\mu,\nu}\varepsilon_{\mu\nu}
    n_{1,\mu}(x)\partial_{\nu}\phi_{2}(x+\hat{\mu})
    =
    \frac{\theta}{2\pi}\sum_{x,\mu,\nu}\phi_{2}(x+\hat{1}+\hat{2})\varepsilon_{\mu\nu}
    \partial_{\mu}n_{1,\nu}(x),
\end{align}
which can be absorbed in the Lagrange-multiplier term by $\tilde{\phi}_1(\tilde{x})\mapsto \tilde{\phi}_1(\tilde{x})-\frac{\theta}{2\pi}\phi_2(x+\hat{1}+\hat{2})$. 
Equivalently, the equation of motion for $\tilde{\phi}_{1}$ imposes $\varepsilon_{\mu\nu}\partial_\mu n_{1,\nu}(x)=0$, and this term vanishes. 
Similarly, the third term can be rewritten as
\begin{align}
    \frac{\theta}{2\pi}\sum_{x,\mu,\nu}\varepsilon_{\mu\nu}
    \partial_{\mu}\phi_{1}(x)n_{2,\nu}(x+\hat{\mu})
    =
    -\frac{\theta}{2\pi}\sum_{x,\mu,\nu}\phi_{1}(x)\varepsilon_{\mu\nu}\partial_{\mu}n_{2,\nu}(x),
\end{align}
which can be absorbed by $\tilde{\phi}_2(\tilde{x})\mapsto \tilde{\phi}_2(\tilde{x})+\frac{\theta}{2\pi}\phi_1(x)$. 
Therefore, we can confirm that $Q_{\text{top}}$ takes integral values when the vortex operators are completely absent.

Now, we can discuss how the equivalence at $\theta$ and $\theta+2\pi$ is realized on the lattice. 
In the continuum, we have seen that the theories at $\theta$ and $\theta+2\pi$ are not the same on the nose but they are related by the nontrivial unitary equivalence, which is summarized in the relation~\eqref{correlation_theta-theta+2pi_continuum}, and this turns out to be true also for the lattice. 
The Boltzmann weight of the path integral at $\theta+2\pi$ is given by 
\begin{equation}
    \exp(-S_{\mathrm{free},\, \theta+2\pi})=\exp(-S_{\mathrm{free},\, \theta}-2\pi \im Q_{\mathrm{top}}). 
\end{equation}
If $Q_{\mathrm{top}}$ were manifestly integer quantized, this gives $\exp(-S_{\theta})$ and the lattice theories at $\theta$ and $\theta+2\pi$ would have become identical in the trivial manner. 
However, this is not the case. As we have seen in the above discussion, we need to absorb non-integral mixed terms (i.e. the second and third terms in \eqref{eq:Qtop_expansion}) by shifting the dual variables, 
\begin{equation}
    \tilde{\phi}_1(\tilde{x})\mapsto \tilde{\phi}_1(\tilde{x}) - \phi_2(x+\hat{1}+\hat{2}),\quad 
    \tilde{\phi}_2(\tilde{x})\mapsto \tilde{\phi}_2(\tilde{x}) + \phi_1(x). 
\end{equation}
This transformation derives the relation, 
\begin{equation}
    \left\{\begin{array}{c}
        \exp\Bigl(\im \tilde{\phi}_1(\tilde{x})\Bigr) \\
        \exp\Bigl(\im \tilde{\phi}_2(\tilde{x})\Bigr)
    \end{array}
    \right\} \text{ at $\theta+2\pi$} 
    \longleftrightarrow 
    \left\{\begin{array}{c}
        \exp\Bigl(\im (\tilde{\phi}_1(\tilde{x})-\phi_2(x+\hat{1}+\hat{2}))\Bigr) \\
        \exp\Bigl(\im (\tilde{\phi}_2(\tilde{x})+ \phi_1(x))\Bigr)
    \end{array}\right\} \text{ at $\theta$}, 
    \label{eq:correlation_theta-theta+2pi_lattice}
\end{equation}
which is nothing but the lattice counter part of \eqref{correlation_theta-theta+2pi_continuum}. 
We note that these relations \eqref{eq:correlation_theta-theta+2pi_lattice} strictly determine the placement of the induced electric charge in the dyon operators on the lattice, which is remarkable and becomes possible thanks to the nice topological property of the modified Villain lattice formulation. 

One can also check that this lattice model reproduces the 't~Hooft anomalies of the continuum theory. The lattice construction of these anomalies is done in Ref.~\cite{Abe:2023uan} based on the Wilson-type formulation with the admissibility condition.

\subsubsection{Exact realization of the strong-weak duality on the lattice}
\label{sec:exact_duality_free_lattice}

The modified Villain formulation proves the exact strong-weak (or electric-magnetic) duality\footnote{
In the context of the $2$d compact boson, this is commonly called the $T$-duality as it comes out of the toroidal compactification of the closed string theory. However, in the connection with the $4$d Maxwell theory, this corresponds to the $S$-transformation and the name of the $T$-transformation is assigned for the different operation, $\theta\mapsto \theta+2\pi$. To circumvent the potential confusion due to this unfortunate situation, we will simply call it the strong-weak duality, instead of calling it as the $S$ or $T$ duality, in this paper. } on the lattice as demonstrated in Refs.~\cite{Sulejmanpasic:2019ytl, Anosova:2022cjm}. 
Combined with the $\theta$ periodicity, the electric-magnetic duality plays the crucial role in the proposal of the Cardy-Rabinovici phase diagram~\cite{Cardy:1981fd}. 
Therefore, let us check its validity for our lattice setup of the $2$d compact boson.

We note that the lattice action~\eqref{eq:latticeVillainAction_free} is quadratic in terms of the $\mathbb{Z}$-valued link variables, $n_{a,\mu}(x)$. 
We can obtain the electric-magnetic duality by using the Poisson summation formula for its summation, $\sum_{n_{a,\mu}}$. 
Appendix \ref{appendix_C} shows the detailed computation of the Poisson summation, which introduces the dual integer-valued gauge field $\tilde{n}_{a,\mu}(\tilde{x})$ living on the links of the dual lattice. 
The partition function~\eqref{eq:latticeZ_free} is then rewritten as 
\begin{align}
    Z_{\mathrm{free}}
    =
    \mathcal{N} &\int\Diff\phi_{a}\Diff\tilde{\phi}_{a}
    \prod_{\tilde{x},\mu}\sum_{\tilde{n}_{a,\mu}(\tilde{x})\in \mathbb{Z}} \exp(-\tilde{S}_{\mathrm{free}}[\phi_a,\tilde{\phi}_a,\tilde{n}_{a,\mu}]), 
    \label{dualization-PF}
\end{align}
where $\mathcal{N}$ is a numerical factor and the dual action $\tilde{S}$ is given by 
\begin{align}
    \tilde{S}_{\mathrm{free}}&=
    \sum_{a}\sum_{\tilde{x},\mu}  \frac{\tilde{R}_{a}^2}{4\pi}[\partial_{\mu}\tilde{\phi}_{a}(\tilde{x}) + 2\pi \tilde{n}_{a,\mu}(\tilde{x})]^{2}
    +\im \sum_{a}\sum_{\tilde{x},\mu,\nu}\varepsilon_{\mu\nu}\phi_{a}(x) \partial_{\mu} \tilde{n}_{a,\nu}(\tilde{x}-\hat{1}-\hat{2})\notag\\
    &\quad+ \im \tilde{\theta} Q_{\text{top}}[\tilde{\phi}_a,\tilde{n}_{a,\mu}]. 
\end{align}
Thus, the dual action $\tilde{S}_{\mathrm{free}}$ takes the same functional form with the original action $S_{\mathrm{free}}$ via
\begin{equation}
    (\phi_a(x), \tilde{\phi}_a(\tilde{x})) \mapsto (\tilde{\phi}_a(\tilde{x}), \phi_a(x+\hat{1}+\hat{2})),
\end{equation}
while the coupling constants are replaced as
\begin{align}
   \tilde{R}_{1}^2 
   = \frac{R_2^2}{R_1^2 R_2^2+ \frac{\theta^{2}}{(2\pi)^{2}}},
   \quad 
   \tilde{R}_{2}^2 
   = \frac{R_1^2}{R_1^2 R_2^2+ \frac{\theta^{2}}{(2\pi)^{2}}} , \quad
   \tilde{\theta} 
   = 
   -\frac{\theta}{R_1^2 R_2^2+ \frac{\theta^{2}}{(2\pi)^{2}}}. 
   \label{new-Coupling}
\end{align} 
This is exactly the same relation with the one obtained by the $2$d Abelian duality in the continuum theory. 
When $R_1^2=R_2^2=R^2$, it is convenient to introduce the holomorphic coupling, 
\begin{equation}
    \tau=\frac{\theta}{2\pi}+\im R^2, 
\end{equation}
and then the duality transformation can be summarized as 
\begin{equation}
    \tau\mapsto \tilde{\tau}=-\tau^{-1}. 
\end{equation}
With the dual coupling constants, the scaling dimension of the operator~\eqref{scaling-dim} can be written in the following way, 
\begin{align}
    &\quad \Delta[\rme^{\im(N_1\phi_1+N_2\phi_2+\tilde{N}_1\tilde{\phi}_1+\tilde{N}_2\tilde{\phi}_2)}]\notag\\
    &= \frac{1}{2 R_1^2}\left(N_{1}+\frac{\theta}{2\pi}\tilde{N}_{2}\right)^{2} + \frac{R_1^2}{2}\tilde{N}_{1}^{2} + \frac{1}{2R_2^2}\left(N_{2}-\frac{\theta}{2\pi}\tilde{N}_{1}\right)^{2} + \frac{R_2^2}{2}\tilde{N}_{2}^{2} \notag\\
    &= \frac{1}{2 \tilde{R}_{1}^2}\left(\tilde{N}_{1}+\frac{\tilde{\theta} }{2\pi}N_{2}\right)^{2}+ \frac{\tilde{R}_1^2}{2}N_{1}^{2} + \frac{1}{2\tilde{R}_{2}^2}\left(\tilde{N}_{2}-\frac{\tilde{\theta} }{2\pi} N_{1}\right)^{2} + \frac{\tilde{R}_2^2}{2} N_{2}^{2}. 
\end{align}
For the case of $R_1^2=R_2^2=R^2$, this expression becomes $\frac{1}{2\mathrm{Im(\tau)}}\left[|N_1+\tilde{N}_2\tau|^2+|N_2-\tilde{N}_1\tau|^2\right]$. 
We can confirm that the scaling dimensions of the vertex and vortex operators are exchanged under the strong-weak duality.

\section{2d Cardy-Rabinovici model and the phase diagram}
\label{sec:{2d Cardy-Rabinovici model with modified Villain lattice formulation}}

In this section, we discuss the reformulation of the $2$d Cardy-Rabinovici model based on the modified Villain lattice. 
In addition, we conjecture the phase diagram via the scaling dimension argument.

\subsection{2d Cardy-Rabinovici model with the modified Villain lattice}
\label{subsec:{Construction of 2d Cardy-Rabinovici model with Villain lattice}}

As mentioned in the Introduction, the $2$d version of the Cardy-Rabinovici (CR) model was proposed already in the original paper by Cardy and Rabinovici~\cite{Cardy:1981qy} based on the dimensional reduction of their $4$d model. 
Since their formulation is based on the conventional Villain lattice, the $\theta$ periodicity and the strong-weak duality are both missing as the lattice theory, while its phase diagram is conjectured based on these two properties. 
Using the modified Villain formulation, we have seen that both properties can be manifestly realized for the case of the $2$d free compact boson in Section~\ref{subsec:{Lattice description of the theta angle and the Witten effect}}, so let us discuss if it can be extended by including the interaction term for the CR model.  
In the following of the discussion, let us set $R_1^2=R_2^2=R^2$ for simplicity. 

For the construction of the $2$d CR model below, we introduce a specific local interaction $V[N \phi_a,\tilde{\phi}_a]$ to the free-theory lattice action~\eqref{eq:latticeVillainAction_free}, 
\begin{equation}
    S_{\mathrm{CR},\, \bar{\theta}}[\phi_a,\tilde{\phi}_a,n_{a,\mu}]
    =S_{\mathrm{free},\, N\bar{\theta}}[\phi_a,\tilde{\phi}_a,n_{a,\mu}] + V[N \phi_a,\tilde{\phi}_a],  
    \label{eq:2dCR_action}
\end{equation} 
where the $\theta$ angle is rescaled by $N$ as $\theta=N\bar{\theta}$. 
Before going into the details, let us briefly mention the criterion/motivation for introducing the specific interaction. 
The original idea of the CR model~\cite{Cardy:1981qy} is to give a simplified model that is helpful to understand behaviors of $4$d $SU(N)$ Yang-Mills theory potentially with adjoint matters. 
Thus, we would like to construct the model that shares the same global symmetry with the $4$d $SU(N)$ Yang-Mills theory, i.e. the $\mathbb{Z}_N$ $1$-form symmetry and the $\theta$ periodicity. 
The $4$d Maxwell theory has the electric and magnetic $U(1)$ symmetries, which is genuinely larger than the one of $SU(N)$ Yang-Mills theory, so we would like to explicitly break it as $U(1)_{\mathrm{ele}}\times U(1)_{\mathrm{mag}}\to (\mathbb{Z}_N)_{\mathrm{ele}}\times \{1\}_{\mathrm{mag}}$. 
To this end, there should be electric charge-$N$ matters and magnetic charge-$1$ matters as dynamical excitations. 
The requirement of the (rescaled) $\theta$ periodicity and the strong-weak duality gives a further constraint on the dynamics of electric/magnetic matters. 
We should add the local potential realizing this idea for the $2$d CR model.

As the local potential term satisfying the above criterion, let us consider the following, 
\begin{align}
    V
    =
    -g\sum_x \sum_{\gcd(p,q)=1}\left[\cos\bigl(Np \phi_1(x)+q \tilde{\phi}_2(\tilde{x})\bigr)+\cos\bigl(Np \phi_2(x)-q \tilde{\phi}_1(\tilde{x}-\hat{1}-\hat{2})\bigr)\right], 
    \label{matter-action_2dC-R}
\end{align}
where $g$ is the coupling constant. Here, the summation over charges, $\sum_{\gcd(p,q)=1}$, is taken over $p,q\in \mathbb{Z}$ that are coprime with one another.\footnote{
The coprime condition is introduced to choose the minimal set of charges invariant under the $\theta$ periodicity and the strong-weak duality. One may replace the sum by $\sum_{p,q\in \mathbb{Z}}$ without the minimality requirement. },\footnote{
The summation over the charges is an infinite sum, so it is a nontrivial question if this lattice theory is really well-defined. In this paper, we assume its well-definedness and give a formal analysis on its symmetry, duality, and phase diagram. 
We may introduce a convergent factor for $\sum_{p,q}$ to avoid potential problems, and we shall see in Sec.~\ref{subsec:{Phase diagram via the scaling dimension argument}}
that this can be achieved with respecting the $2\pi$ periodicity of $\bar{\theta}$ and also the strong-weak duality. } 
This potential respects the $\mathbb{Z}_N\times \mathbb{Z}_N$ electric symmetry generated by $\phi_a\mapsto \phi_a+\frac{2\pi}{N}$, while the magnetic $U(1)$ symmetry is completely broken: 
\begin{equation}
    (\mathbb{Z}_N\times \mathbb{Z}_N)_{\mathrm{ele}}\times (\{1\}\times \{1\})_{\mathrm{mag}}\subset (U(1)\times U(1))_{\mathrm{ele}}\times (U(1)\times U(1))_{\mathrm{mag}}. 
\end{equation}
These $\mathbb{Z}_N$ symmetries are the counterpart of the $\mathbb{Z}_N$ center symmetries for the $4$d gauge theory. 
We also note that the all vertex/vortex operators in the potential have the spin $0$, which is consistent with the Euclidean rotational invariance.\footnote{As noted in the last paragraph of Sec.~\ref{subsubsec:{Construction of the modified Villain lattice action}}, the Euclidean $\frac{\pi}{2}$ rotation with the lattice topological charge is given by the $\frac{\pi}{2}$ rotation around the site $x_*$ for the $\phi_1$ field and the dual site $\tilde{x}_*=x_*+\frac{1}{2}(\hat{1}+\hat{2})$ for the $\phi_2$ field. Thus, the Euclidean rotational invariance becomes more manifest if we shift the location of $\phi_2,\tilde{\phi}_2$ by $-\frac{1}{2}(\hat{1}+\hat{2})$, and each term in the potential becomes on-site after this redefinition. See also Appendix~\ref{appendix_B}. } 
Let us check if this model actually enjoys the rescaled $\theta$ periodicity and also the strong-weak duality. 

\subsubsection{Rescaled \texorpdfstring{$\theta$}{theta} periodicity and its mixed 't~Hooft anomaly}

In this subsection, let us show that the $2$d CR model defined by \eqref{eq:2dCR_action} and \eqref{matter-action_2dC-R} enjoys the rescaled $\theta$ periodicity, $\bar{\theta}\sim \bar{\theta}+2\pi$ with $\theta=N\bar{\theta}$, and we also discuss its generalized mixed 't~Hooft anomaly with the electric $\mathbb{Z}_N\times \mathbb{Z}_N$ symmetry. 

To see the equivalence between $\bar{\theta}$ and $\bar{\theta}+2\pi$, we explicitly write down the Boltzmann weight for the $2$d CR model at $\bar{\theta}+2\pi$ as
\begin{align}
    &\quad \exp(-S_{\mathrm{CR},\, \bar{\theta}+2\pi}[\phi_a,\tilde{\phi}_a,n_{a,\mu}]) \notag\\ 
    &= \exp(-S_{\mathrm{free},\, N(\bar{\theta}+2\pi)}[\phi_a,\tilde{\phi}_a,n_{a,\mu}]-V[N\phi_a,\tilde{\phi}_a])\notag\\
    &= \exp(-S_{\mathrm{free},\, N\bar{\theta}}[\phi_a,\tilde{\phi}_a,n_{a,\mu}]-V[N\phi_a,\tilde{\phi}_a] -2\pi \im N Q_{\mathrm{top}}). 
\end{align}
The integer-quantized part of $Q_{\mathrm{top}}$ automatically drops in $\exp(-2\pi \im N Q_{\mathrm{top}})$, so we need to absorb its non-integer-quantized part by shifting the dual fields $\tilde{\phi}_a$ as we have seen in Sec.~\ref{subsubsec:{Exact realization of the Witten effect on the lattice}}: 
\begin{equation}
    \tilde{\phi}_{1}(\tilde{x})
    \rightarrow
    \tilde{\phi}_{1}(\tilde{x}) - N\phi_{2}(x+\hat{1}+\hat{2}),\quad 
    \tilde{\phi}_{2}(\tilde{x}) 
    \rightarrow
    \tilde{\phi}_{2}(\tilde{x}) + N\phi_{1}(x).
\end{equation}
Thus, the equivalence between $\bar{\theta}$ and $\bar{\theta}+2\pi$ is established if and only if the local potential satisfies 
\begin{equation}
    V\bigl[N\phi_a(x), \tilde{\phi}_{1}(\tilde{x}) - N\phi_{2}(x+\hat{1}+\hat{2}), \tilde{\phi}_{2}(\tilde{x}) + N\phi_{1}(x)\bigr] = V\bigl[N\phi_a(x), \tilde{\phi}_1(\tilde{x}), \tilde{\phi}_2(\tilde{x})\bigr]. 
\end{equation}
This property is satisfied for \eqref{matter-action_2dC-R} by shifting $p$ by $p-q$ in the sum for the charges since $\gcd(p-q,q)=\gcd(p,q)$. 

This argument clarifies why we need the rescaling of the $\theta$ angle as $\theta=N\bar{\theta}$ for its $2\pi$ periodicity~\cite{Honda:2020txe, Hayashi:2022fkw}. 
In the CR model, the electric charge in the Lagrangian is quantized in $N$ while the magnetic charge takes general integers. Without the rescaling, the $2\pi$ shift of the $\theta$ angle causes the Witten effect $\tilde{\phi}_a \mapsto \tilde{\phi}_a - \varepsilon_{ab}\phi_b$, and the induced electric charge is not quantized in $N$, which changes the shape of the local potential. 
The minimal periodicity for the unitary equivalence is then given by $2\pi N$ for the non-rescaled $\theta$ angle. 

Let us then calculate the generalized mixed 't~Hooft anomaly between the $\mathbb{Z}_N\times \mathbb{Z}_N$ symmetry and the rescaled $\theta$ periodicity, $\bar{\theta}\sim \bar{\theta}+2\pi$. 
To this end, we introduce the flat background gauge field $A_{a}$ for the $\mathbb{Z}_N\times \mathbb{Z}_N$ symmetry, and then we shall obtain that 
\begin{equation}
    Z_{\mathrm{CR},\bar{\theta}+2\pi}[A_a]=\exp\left(-\frac{2\pi \im}{N}\int A_1\cup A_2\right)Z_{\mathrm{CR},\bar{\theta}}[A_a],
    \label{eq:generalized_anomaly_theta}
\end{equation}
where $Z_{\mathrm{CR},\bar{\theta}}[A_a]$ is the partition function of the $2$d CR model coupled to the background gauge field and 
\begin{equation}
    \int A_1\cup A_2=\sum_{x,\mu\nu}\varepsilon_{\mu\nu}A_{1,\mu}(x)A_{2,\nu}(x+\hat{\mu}).
\end{equation} 
Here, the $\mathbb{Z}_N$ gauge field is realized as the $\mathbb{Z}$-valued link variables $A_{a,\mu}(x)$ with the mod-$N$ identification and we also impose the flatness condition\footnote{Precisely speaking, the flatness condition for the $\mathbb{Z}_N$ gauge field only requires $\varepsilon_{\mu\nu}\partial_\mu A_{a,\nu}(x)=0$ in mod $N$ when $A_{a,\mu}(x)$ is lifted to an integer-valued cochain. However, for the torsion-free manifolds such as sphere, torus, etc., we may take the lift to an integer-valued cocycle thanks to the Bockstein sequence. } 
\begin{align}
    \sum_{\mu\nu}\varepsilon_{\mu\nu}\partial_{\mu}A_{a,\nu}(x)
    = 0.
    \label{B_flatness-cond}
\end{align}
The anomalous relation is nothing but the $2$d reduction of the $4$d anomaly for the $SU(N)$ Yang-Mills theory~\cite{Gaiotto:2017yup} and the $4$d CR model~\cite{Honda:2020txe, Hayashi:2022fkw}.

When we introduce the flat $\mathbb{Z}_N\times \mathbb{Z}_N$ background gauge field, we replace the usual lattice derivative $\partial_\mu\phi_a(x)+2\pi n_{a,\mu}(x)$ by the covariant derivative, 
\begin{equation}
    \partial_\mu\phi_a(x)+2\pi n_{a,\mu}(x) \Rightarrow \partial_\mu\phi_a(x)+2\pi n_{a,\mu}(x)+\frac{2\pi}{N}A_{a,\mu}(x), 
\end{equation}
in the action $S_{\mathrm{CR},\, \bar{\theta}}$. 
Then, the electric gauge invariance~\eqref{eq:electric_gauge_trans} is extended as 
\begin{equation}
    \left\{\begin{array}{l}
        \phi_{a}(x)
        \rightarrow
        \phi_{a}(x) - 2\pi k_{a}(x)-\frac{2\pi}{N}l_{a}(x), \\
        n_{a,\mu}(x)
        \rightarrow
        n_{a,\mu}(x) + \partial_{\mu}k_{a}(x)-m_{a,\mu}(x), \\
        A_{a,\mu}(x)
        \rightarrow
        A_{a,\mu}(x) + \partial_{\mu}l_{a}(x) + Nm_{a,\mu}(x), 
    \end{array}\right.
\end{equation}
where $k_{a},\tilde{k}_{a}$ and $l_{a}$ are zero-form integer gauge transformation parameters, and $m_{a}$ are one-form integer gauge transformation parameters satisfying $\varepsilon_{\mu\nu}\partial_{\mu}m_{a,\nu}(x)=0$. 
The invariance for the kinetic term and $Q_{\mathrm{top}}$ would be obvious as the covariant derivative is manifestly invariant under these transformation. 
The invariance for the Lagrange-multiplier term might be less obvious, but it becomes manifest if we rewrite it as $\tilde{\phi}_{a}(\tilde{x}) \varepsilon_{\mu\nu}\partial_{\mu} n_{a,\nu}(x)=\frac{1}{2\pi}\tilde{\phi}_{a}(\tilde{x}) \varepsilon_{\mu\nu}\partial_{\mu} [\partial_\nu\phi_a(x)+2\pi n_{a,\nu}(x)+\frac{2\pi}{N}A_{a,\nu}(x)]$;\footnote{
When we realize the flat $\mathbb{Z}_N$ gauge field as a lift to the integer-valued $1$-cocycle, these two expressions are completely identical. But, if we realize the $\mathbb{Z}_N$ gauge field as an integer-valued cochain that is closed only in mod $N$, then we must use the second expression for the gauge invariance. In this sense, one may claim the second expression is more essential. } the first term on the right-hand-side vanishes automatically and the last term also vanishes due to the flatness of $A_{a}$.  

To derive the 't~Hooft anomalous relation, we need to relate $\exp(-S_{\mathrm{CR},\, \bar{\theta}+2\pi})$ and $\exp(-S_{\mathrm{CR},\, \bar{\theta}})$ under the presence of the background gauge field, so let us evaluate
\begin{align}
    \exp(-2\pi \im N Q_{\mathrm{top}})
    &=\exp\left[-\frac{\im N}{2\pi}\int \left(\diff \phi_1+2\pi n_1+\frac{2\pi}{N}A_1\right)\cup \left(\diff \phi_2+2\pi n_2+\frac{2\pi}{N}A_2\right)\right]\notag \\
    &= \exp\left[\im \int\left(N\phi_1\cup \diff n_2-\diff n_1\cup N\phi_2-\frac{2\pi}{N}A_1\cup A_2\right)\right]. 
\end{align}
The first two terms are absorbed by redefining $\tilde{\phi}_{1,2}$, and we obtain the extra phase factor that depends only on the background gauge field, which gives \eqref{eq:generalized_anomaly_theta}.

In addition, let us see the parity symmetry $\mathsf{P}$ (or $\mathsf{CP}$) in this model. Like the Euclidean $\frac{\pi}{2}$ rotation, $Q_{\mathrm{top}}$ does not have a nice transformation under the naive parity operation. 
If we apply the parity for $\phi_1, \tilde{\phi}_1$ about the original lattice and for $\phi_2,\tilde{\phi}_2$ about the dual lattice, we can check that the lattice topological charge $Q_{\mathrm{top}}$ flips its sign as expected in the continuum formulation, and we call it as $\mathsf{P}$ on the lattice. 
Then, we find  
\begin{align}
    \mathsf{P}: 
    \bar{\theta}
    \rightarrow
    -\bar{\theta}. 
\end{align}
Due to its $2\pi$ periodicity, $\mathsf{P}$ is a good symmetry when $\bar{\theta}$ is quantized in $\pi$. In particular, $\bar{\theta}=\pi$ is a nontrivial $\mathsf{P}$-symmetric point of the $2$d CR model. 
This parity is potentially 't~Hooft anomalous: 
\begin{align}
    \mathsf{P}: Z_{\mathrm{CR},\, \bar{\theta}=\pi}[A_a]\mapsto Z_{\mathrm{CR},\, \bar{\theta}=-\pi}[A_a]=\rme^{\frac{2\pi \im}{N}\int A_1\cup A_2} Z_{\mathrm{CR},\, \bar{\theta}=\pi}[A_a]. 
    \label{eq:CP-ZN_anomaly}
\end{align}
To see if this is a genuine anomaly, we need to check that any local counterterms of the background gauge field cannot eliminate it. 
The possible local counterterm is given by 
\begin{equation}
    Z^{(k)}_{\mathrm{CR},\, \bar{\theta}}[A_a]=\exp\left(\frac{2\pi \im k}{N}\int A_1\cup A_2\right)Z_{\mathrm{CR},\, \bar{\theta}}[A_a], 
\end{equation}
where $k\in \mathbb{Z}$ for the gauge invariance and $k\sim k+N$. Since this local counterterm also flips its sign under $\mathsf{P}$, we find that 
\begin{equation}
    \mathsf{P}: Z^{(k)}_{\mathrm{CR},\, \bar{\theta}=\pi}[A_a]\mapsto \exp\left(-\frac{2\pi \im (2k-1)}{N}\int A_1\cup A_2\right) Z^{(k)}_{\mathrm{CR},\, \bar{\theta}=\pi}[A_a]. 
\end{equation}
Thus, the potentially anomalous phase is canceled by the local counterterm if and only if there exists $k$ such that 
\begin{align}
    2k-1 
    =
    0 \quad (\text{mod}\; N).
\end{align}
When $N$ is even, such $k$ does not exist, and we have the mixed 't~Hooft anomaly between $\mathbb{Z}_N\times \mathbb{Z}_N$ and $\mathsf{P}$ that constrains the low-energy physics: The low-energy physics must spontaneously break the symmetry down to an anomaly-free subgroup if it is gapped or must support gapless excitations. 
For odd $N$, $k=\frac{N+1}{2}$ cancels the anomaly, and we need to discuss the global inconsistency between the physics at $\bar{\theta}=0$ and $\bar{\theta}=\pi$ instead of the physics at $\bar{\theta}=\pi$ itself~\cite{Gaiotto:2017yup, Tanizaki:2017bam, Kikuchi:2017pcp,  Komargodski:2017dmc, Tanizaki:2018xto, Cordova:2019jnf}: If the low-energy physics at $\bar{\theta}=\pi$ is trivially gapped, it has to be nontrivial as a symmetry-protected topological (SPT) state compared with the one at $\bar{\theta}=0$.

\subsubsection{Strong-weak duality of the 2d CR model with the \texorpdfstring{$\mathbb{Z}_N\times \mathbb{Z}_N$}{ZNxZN} gauging}

In Sec.~\ref{sec:exact_duality_free_lattice}, we have confirmed that the strong-weak duality is exact for the free theory on the lattice in the modified Villain formulation. 
This duality is derived by the Poisson summation for $n_{a,\mu}$, so the derivation is applicable with the potential term as $V$ does not contain $n_{a,\mu}$. 
However, the strong-weak duality in Sec.~\ref{sec:exact_duality_free_lattice} exchanges $\phi_a \leftrightarrow \tilde{\phi}_a$, which is not respected for the local potential $V$ of the $2$d CR model: The charges of $\phi_a$ are quantized in $N$ while the ones for $\tilde{\phi}_a$ are not. 
Thus, we need another strong-weak duality that exchanges $N\phi_a$ and $\tilde{\phi}_a$ to apply it for the $2$d CR model.

Such a strong-weak duality is obtained by 
\begin{itemize}
    \item[(1)] gauging the electric $\mathbb{Z}_N\times \mathbb{Z}_N$ symmetry and then  
    \item[(2)] applying the strong-weak duality in Sec.~\ref{sec:exact_duality_free_lattice}. 
\end{itemize}
To discuss the 't~Hooft anomaly, we introduced the background gauge field $A_{a}$ for the $\mathbb{Z}_N\times \mathbb{Z}_N$ symmetry. When gauging $\mathbb{Z}_N\times \mathbb{Z}_N$, we promote $A_a$ as the dynamical field. We note that $A_{a,\mu}$ always appears in the combination of $N n_{a,\mu}+A_{a,\mu}$, and we may regard it as the new $\mathbb{Z}$-valued link variable, 
\begin{equation}
    n'_{a,\mu}(x):=N n_{a,\mu}(x) + A_{a,\mu}(x). 
\end{equation}
Then, the gauged free action becomes 
\begin{align}
    S_{\text{gauged free},\, N\bar{\theta}} &= \sum_{a}\sum_{x,\mu}\frac{R^2/N^2}{4\pi}(\partial_{\mu}(N \phi_a)+2\pi n'_{a,\mu})^2+\frac{\im N\bar{\theta}}{N^2}Q_{\mathrm{top}}[N\phi_a, n'_{a,\mu}] \notag\\
    &\quad +\im \sum_{a}\sum_{x,\mu\nu}\left(\frac{1}{N}\tilde{\phi}_{a}(\tilde{x})\right) \varepsilon_{\mu\nu} \partial_{\mu}n'_{a,\nu}(x), 
\end{align}
and we obtain the same functional form of the free-theory action by introducing 
\begin{equation}
    \phi'_a(x) = N\phi_a(x),\quad \tilde{\phi}'_{a}(\tilde{x})=\frac{1}{N}\tilde{\phi}_a(\tilde{x}). 
\end{equation}
Thus, the strong-weak duality operation after the $\mathbb{Z}_N\times \mathbb{Z}_N$ gauging gives the following transformations, 
\begin{equation}
    \left\{\begin{array}{c}
         \exp(\im N \phi_a) \\
         \exp(\im \tilde{\phi}_a) \\
         \tau
    \end{array}\right\} \xrightarrow{\text{$\mathbb{Z}_N\times \mathbb{Z}_N$ gauging} }
    \left\{\begin{array}{c}
         \exp(\im \phi_a) \\
         \exp(\im N \tilde{\phi}_a) \\
         \tau/N^2
    \end{array}\right\} \xrightarrow{\text{strong-weak duality}}
    \left\{\begin{array}{c}
         \exp(\im \tilde{\phi}_a) \\
         \exp(\im N  \phi_a) \\
         -N^2/\tau
    \end{array}\right\},
\end{equation}
where $\tau=\frac{\theta}{2\pi}+\im R^2=N\left(\frac{\bar{\theta}}{2\pi}+\frac{\im}{N}R^2\right)$. 
It is straightforward to confirm that the potential \eqref{matter-action_2dC-R} of the $2$d CR model $V$ is invariant under this duality operation by exchanging $p$ and $-q$, so this is the correct strong-weak duality operation for the $2$d CR model. 
The duality ensures that the physics at $\tau$ and $\tau'=-N^2/\tau$ becomes identical up to the $\mathbb{Z}_N\times \mathbb{Z}_N$ gauging operation. 

At the self-dual point, such as $\tau/N=\im$, the strong-weak duality gives an extra symmetry for the $2$d CR model. 
The fact that the duality operation is associated with the $\mathbb{Z}_N\times \mathbb{Z}_N$ gauging makes the duality symmetry non-invertible~\cite{Aasen:2016dop, Aasen:2020jwb, Koide:2021zxj, Kaidi:2021xfk, Choi:2021kmx}: For example, the fusion of the duality symmetry operators at $\tau/N=\im$ gives the condensation defect for the $\mathbb{Z}_N\times \mathbb{Z}_N$ symmetry. When the duality symmetry operator surrounds $\exp(\im \phi_a)$, it gives $0$ and thus the duality symmetry operator is non-invertible.  

\subsection{Phase diagram based on the perturbative renormalization group}
\label{subsec:{Phase diagram via the scaling dimension argument}}

In this section, let us determine the phase diagram when $g\ll 1$ at the lattice scale and then the effect of $V$ can be treated as the perturbation. 
In this case, the phase diagram can be determined by the perturbative renormalization group based on the simple scaling dimension argument, following Refs.~\cite{Cardy:1981qy,Cardy:1981fd}.

The scaling dimension for $\cos(Np \phi_1+q\tilde{\phi}_2)$ and $\cos(Np\phi_2-q\tilde{\phi}_1)$ is given by 
\begin{align}
    \Delta_{p,q} & := \Delta[\cos(Np \phi_1+q\tilde{\phi}_2)]=\Delta[\cos(Np\phi_2-q\tilde{\phi}_1)]\notag\\
    &=\frac{N}{2\mathrm{Im(\tau/N)}}\left|p+\frac{\tau}{N}q\right|^2. 
\end{align}
Let $E$ be the renormalization energy scale and we write the local potential at the scale $E$ as 
\begin{equation}
    V=-\int \diff^2 x \sum_{\gcd(p,q)=1} g_{p,q}(E)[\cos(Np \phi_1+q \tilde{\phi}_2)+ \cos(Np \phi_2-q\tilde{\phi}_1)]. 
\end{equation}
Let us introduce the dimensionless couplings $\bar{g}_{p,q}(E)=g_{p,q}(E)/E^2$, and then $\bar{g}_{p,q}=g$ for all $p,q$ at the lattice energy scale, which gives the initial condition for the renormalization group. The perturbative renormalization group at the leading order is given by
\begin{equation}
    \frac{\diff \bar{g}_{p,q}}{\diff \ln E}=-(2-\Delta_{p,q})\bar{g}_{p,q},  
\end{equation}
and the dimensionless coupling shows the scaling behavior,  
\begin{equation}
    \bar{g}_{p,q}(E)\sim (1/E)^{2-\Delta_{p,q}} 
    \to \left\{
    \begin{array}{cc}
       +\infty  &  (\text{as $E\to 0$ for $\Delta_{p,q}<2$}), \\
       0  & (\text{as $E\to 0$ for $\Delta_{p,q}>2$}). 
    \end{array}
    \right. 
\end{equation}
Thus, the running couplings depend on $p,q$ and also on $\tau$ via the scaling dimension $\Delta_{p,q}$, and the infinite sum over $p,q$ becomes convergent. 
We may introduce such $(p,q,\tau)$-dependent lattice couplings from the beginning when we define the lattice potential in \eqref{matter-action_2dC-R}, and then we can define the manifestly finite lattice theory that enjoys the exact $\bar{\theta}$ periodicity and strong-weak duality.\footnote{We thank Tin Sulejmanpasic for pointing it out. } 

We then identify the low-energy physics of the $2$d CR model by the following rule: 
\begin{enumerate}
    \item When $\Delta_{p,q}>2$ for any $p,q$, all the couplings $\bar{g}_{p,q}$ become zero in the deep infrared, and we obtain the free compact boson. 
    Thus, the system is gapless, which is a counterpart of the $4$d Coulomb phase. 
    \item When $\Delta_{p,q}<2$ for some $p,q$, the mass gap is generated by the relevant perturbations, $\cos(Np \phi_1+q\tilde{\phi}_2)$ and $\cos(N p \phi_2-q\tilde{\phi}_1)$. 
    When there are several $p,q$ such that $\Delta_{p,q}<2$, we assume that the mass gap is generated by the most relevant perturbation. 
\end{enumerate}

\begin{figure}
    \centering
    \includegraphics[width=0.75\linewidth]{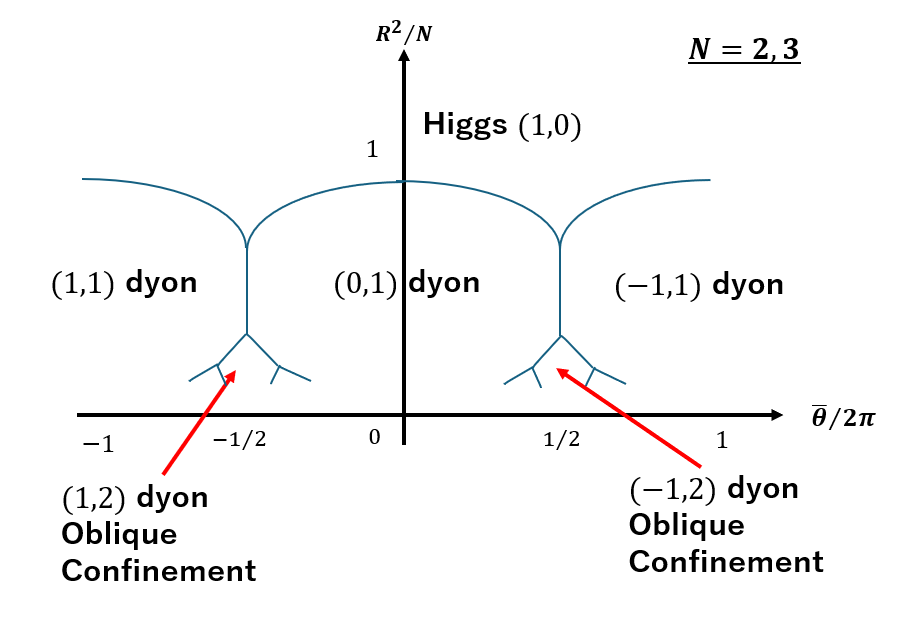}
    \caption{The phase diagram of the reformulated $2$d Cardy-Rabinovici model for $N=2,3$ with $g\ll 1$. The vertical axis denotes the rescaled compact-boson radius $R^2/N$, and the horizontal axis denotes the rescaled vacuum angle, $\frac{\bar{\theta}}{2\pi}=\frac{\theta}{2\pi N}$. 
    The phase diagram is always gapped, and it has the rich structure of quantum phase transitions. 
    For $N\ge 4$, the gapless phase starts to appear. }
    \label{fig:2d CR-phasediagram_N=23}
\end{figure}

When $N<2\sqrt{3}$ (i.e. $N=2$ or $3$), we can check there exist some relevant $p,q$ for any $\tau/N$, and thus the Coulomb phase does not arise at all. 
The phase diagram for sufficiently small $g$ is shown in Fig.~\ref{fig:2d CR-phasediagram_N=23}, and 
the meaning of each gapped phases will be discussed in the next section. 
Although the phase diagram shown in Fig.~\ref{fig:2d CR-phasediagram_N=23} was expected for the original version of the $4$d/$2$d Cardy-Rabinovici model~\cite{Cardy:1981qy, Cardy:1981fd}, the original model does not show this phase diagram in the precise sense due to the violation of the $\theta$ periodicity and the strong-weak duality. 
Let us emphasize that it becomes the exact result for our model with the modified Villain formulation at $g\ll 1$ because of the $\theta$ periodicity and strong-weak duality. 
For $N\ge 4$, the gapless phase appears and it separates the gapped phases. 

\subsection{Gapped phases and the global symmetry}
\label{sec:gapped_phases}

Let $p,q$ be the charge with the minimal scaling dimension, $\Delta_{p,q}<2$, and then the low-energy physics is gapped due to the relevant interactions, $\cos(Np\phi_1+q\tilde{\phi}_2)$ and $\cos(Np \phi_2-q\tilde{\phi}_1)$. 
In this section, we discuss the classification of those gapped phases in Fig.~\ref{fig:2d CR-phasediagram_N=23} from the viewpoint of the $\mathbb{Z}_N\times \mathbb{Z}_N$ symmetry. 

\subsubsection{Higgs phase \texorpdfstring{$(p,q)=(1,0)$}{(p,q)=(1,0)} and spontaneous symmetry breaking}
\label{subsubsec:{Higgs phase}}

Let us consider the case where the gap is generated by $-g\cos (N\phi_{1,2})$, and we call it the Higgs phase as this corresponds to the condensation of the charge-$N$ electric matter in the $4$d context. 
The scaling dimension of the operator is 
\begin{align}
    \Delta_{1,0}
    =
    \frac{N}{2\mathrm{Im}(\tau/N)}=\frac{N^2}{2R^2}, 
\end{align}
which is independent of the $\theta$ angle. 
Therefore, the Higgs phase is the most favored gapped phase when $R^2$ is sufficiently large.

In this phase, the electric symmetry is spontaneously broken,
\begin{align}
    \mathbb{Z}_{N}\times \mathbb{Z}_{N}
    \xrightarrow{\text{SSB}}
    1.
\end{align}
To minimize the relevant perturbation, $-g[\cos(N \phi_1)+\cos(N\phi_2)]$, we need to set $N\phi_1=N\phi_2=0$ mod $2\pi$. Therefore, for some $v>0$, the vacuum expectation values of the vertex operators are given as 
\begin{align}
    \langle \rme^{\im \phi_1}
    \rangle
    = v\,\rme^{\frac{2\pi \im}{N}k_1},\quad 
    \langle \rme^{\im \phi_2}
    \rangle
    = v\,\rme^{\frac{2\pi \im}{N}k_2},
\end{align}
with $k_1,k_2=0,1,\ldots, N-1$. There are $N^2$-fold degenerate ground states due to the spontaneous breaking. 

As the scaling dimension $\Delta_{1,0}$ is independent of the $\theta$ angle, the Higgs phase is realized also at $\bar{\theta}=\pi$ for $R^2\gg 1$. Then, this phase should satisfy the anomaly matching condition of the 't~Hooft anomaly~\eqref{eq:CP-ZN_anomaly}, so let us see how it is satisfied. 
When we turn on a nontrivial background gauge field $A_a$, the phase of the local condensate $\langle \rme^{\im \phi_a}\rangle$ is rotated by the Aharonov–Bohm effect of $A_a$. 
The nontrivial holonomy of $A_a$ requires that the vacua of the different labels $k_a$ should be connected by the domain wall, and thus the partition function should be exponentially suppressed by the domain-wall energy. 
As a result, the low-energy limit of the partition function in the Higgs phase is given by 
\begin{align}
    Z_{\text{Higgs}}[A_{1},A_{2}]
    \propto \delta(A_1)\delta(A_2).
    \label{eq:Higgs_Z}
\end{align}
Since the anomalous phase in \eqref{eq:CP-ZN_anomaly} is nontrivial only if the partition function vanishes, the anomaly matching is satisfied in the low-energy effective theory of the Higgs phase.

Let us understand the behavior~\eqref{eq:Higgs_Z} more explicitly from the lattice model. 
For simplicity, we may send the couplings for $-g\cos{(N\phi_{1,2})}$ to $g=\infty$, and then $\phi_{1,2}$ are restricted to $\frac{2\pi}{N}\mathbb{Z}\subset \mathbb{R}$. 
The minimization of the kinetic term is achieved if 
\begin{align}
    \partial_{\mu}\phi_a(x)+2\pi\left(n_{a,\mu}(x)
    +
    \frac{A_{a,\mu}(x)}{N}\right)
    =
    0, 
\end{align}
which is possible if the holonomy of $A_{a}$ is trivial. When $A_{a}$ has a nontrivial holonomy, there has to be some links that violate the above condition, which causes the domain-wall energy.

\subsubsection{Confinement phases \texorpdfstring{$q=1$}{q=1} as symmetry-protected topological states}
\label{subsubsec:{Confinement phase}}

Let us consider the gapped phases caused by the magnetic condensation, where the most relevant charge is given by $(p,q)=(p,1)$ with some $p$, and we call them confinement phases. 
Its scaling dimension is given by 
\begin{equation}
    \Delta_{p,1}=\frac{N}{2\mathrm{Im}(\tau/N)}|p+\tau/N|^2=\frac{N}{2}\left[\frac{N}{R^2}\left(p+\frac{\bar{\theta}}{2\pi}\right)^2+\frac{R^2}{N}\right]. 
\end{equation}
To achieve the confinement phase with $g\ll 1$, we need small $R^2/N$ to suppress the second term (or the magnetic energy) of the scaling dimension. 
While this makes the electric energy (i.e. the first term of the scaling dimension) larger, it can be suppressed when $|p+\frac{\bar{\theta}}{2\pi}|$ is sufficiently small. 
Thus, the confinement phase strongly depends on the $\theta$ angle.

In the confinement phase, the $\mathbb{Z}_N\times \mathbb{Z}_N$ symmetry is unbroken: $\langle \rme^{\im \phi_1}\rangle=\langle\rme^{\im \phi_2}\rangle=0$. 
Since the relevant perturbation $\cos(Np \phi_1+\tilde{\phi_2})$ and $\cos(Np\phi_2-\tilde{\phi}_1)$ are the vortex operators, the phase of the order parameters $\rme^{\im \phi_a}$ cannot be determined when the vortices are proliferated in the vacuum. 
Thus, the confinement phases are not characterized by the local order parameters. 

To discriminate confinement phases with different $p$, we need to look at the response of the partition function to the background gauge field:
\begin{equation}
    Z_{\text{$(p,1)$ dyon}}[A_a]\propto \exp\left(\frac{2\pi \im p}{N}\int A_1\cup A_2\right). 
    \label{eq:Conf_Z}
\end{equation}
The gauge invariance for the background gauge field requires that the coefficient of the topological term has to be quantized. Thus, this level $p$ cannot jump without quantum phase transitions, and we can discriminate the confinement states with different $p$ mod $N$ as the symmetry-protected topological (SPT) states. 

We can derive this result~\eqref{eq:Conf_Z} starting from the partition function~\eqref{eq:Higgs_Z} of the Higgs phase by applying the strong-weak duality and also the anomalous relation~\eqref{eq:generalized_anomaly_theta} as shown in Ref.~\cite{Hayashi:2022fkw}. 
As the strong-weak duality exchanges $N\phi_a$ and $\tilde{\phi}_a$, we can obtain the $(0,1)$ condensate by applying it to the Higgs phase. 
Since the strong-weak duality of the $2$d CR model involves the gauging of $\mathbb{Z}_N\times \mathbb{Z}_N$, we find 
\begin{equation}
    Z_{\text{$(0,1)$ dyon}}[A_{1,2}]\propto \sum_{[a_i]\in H^1(M_2;\mathbb{Z}_N)} \rme^{\frac{2\pi \im}{N}\int \sum_{i=1,2}A_i\cup a_i}Z_{\mathrm{Higgs}}[a_{1,2}] = 1. 
\end{equation}
The $(p,1)$ condensate is obtained by the Witten effect for the $(0,1)$ condensate via $\bar{\theta}+2\pi p\to \bar{\theta}$, and thus its partition function can be obtained by \eqref{eq:generalized_anomaly_theta} as 
\begin{equation}
    Z_{\text{$(p,1)$ dyon}}[A_{1,2}]\propto \rme^{\frac{2\pi \im p}{N}\int A_1\cup A_2}Z_{\text{$(0,1)$ dyon}}[A_{1,2}]\propto \rme^{\frac{2\pi \im p }{N}\int A_1\cup A_2}, 
\end{equation}
and we obtain \eqref{eq:Conf_Z}.

\begin{figure}
    \centering
    \includegraphics[width=0.75\linewidth]{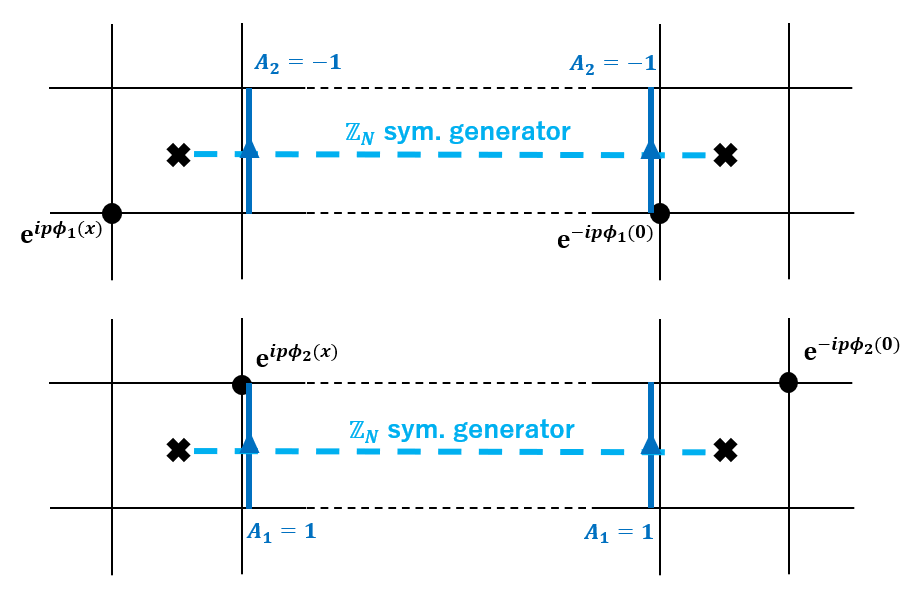}
    \caption{The string order parameters detect the SPT-level $p$ of each confinement phase. 
    The fractional vortices $\rme^{\pm i \frac{1}{N} \tilde{\phi}_{1,2}}$  are non-local operators, but they live on the boundaries for each $\mathbb{Z}_{N}$ symmetry generator, which is realized by introducing the ``non''-flat background $\mathbb{Z}_N$ gauge field $A_{a}$. 
    They have non-vanishing expectation values in the limit $|x|\to \infty$ for a suitable choice of $p$ mod $N$. }
    \label{fig:2dCR_conf_stringorder}
\end{figure}

We can also understand the SPT level $p$ by using the string order parameter~\cite{denNijs:1989ntw, Kennedy:1992tke} (see also Refs.~\cite{Li:2023ani, Maeda:2025ycr} for a recent field-theoretic discussion). 
The relevant perturbation for the $(p,1)$ condensate is given by $\cos(Np \phi_1+\tilde{\phi}_2)$ and $\cos(Np \phi_2-\tilde{\phi_1})$, we may think that their fractional operators 
\begin{equation}
    \rme^{\im \left(p \phi_1+\frac{1}{N}\tilde{\phi}_2\right)},\quad \rme^{\im\left(p\phi_2-\frac{1}{N}\tilde{\phi}_1\right)}
\end{equation}
have nonzero expectation values. 
We should note that $\rme^{\pm \im \frac{1}{N}\tilde{\phi}_{1,2}}$ are nonlocal operators but they live on the boundaries for the $\mathbb{Z}_N\times \mathbb{Z}_N$ symmetry generators, and 
such operators are called string order parameters (see Fig. \ref{fig:2dCR_conf_stringorder}). 
The boundary of the string must be dressed by the $\mathbb{Z}_N\times \mathbb{Z}_N$ charged operators, $\rme^{\im p \phi_{1,2}}$, and this detects the level $p$ of the SPT state. 
This corresponds to the classification of the $4$d confinement states by the Wilson-'t~Hooft classification~\cite{tHooft:1977nqb, tHooft:1979rtg, Nguyen:2023fun}. 

Let us now assume that the confinement states are realized at $\bar{\theta}=\pi$, and then we need to consider how the anomaly~\eqref{eq:CP-ZN_anomaly} is satisfied by the ground states. 
For $g\ll 1$, this situation can be realized only when $N=2,3$ (and possibly $4$), as the system becomes gapless at $\bar{\theta}=\pi$ when $N> 4$ according to the perturbative renormalization group. 
Since the smallest scaling dimensions of $\Delta_{p,1}$ at $\bar{\theta}=\pi$ are given by $p=0$ and $p=-1$, we expect to find the two-fold ground states under this assumption. 
These two vacua are related by $\mathsf{P}$ at $\bar{\theta} = \pi$, that is, the system shows the spontaneous $\mathsf{P}$ breaking, $(\mathbb{Z}_2)_{\mathsf{P}}\xrightarrow{\text{SSB}}1$. 
To confirm the anomaly~\eqref{eq:CP-ZN_anomaly}, let us write down the low-energy partition function of this system, 
\begin{align}
    Z[A_{1,2}]
    &=
    Z_{\text{$(0,1)$ dyon}}[A_{1,2}] + Z_{\text{$(-1,1)$ dyon}}[A_{1,2}]\notag\\
    &\propto 
    1 
    + \exp{\bigg(-
    \frac{2\pi \im}{N} \int A_{1} \cup A_{2}
    \bigg)}.
\end{align}
Under the $\mathsf{P}$ (or $\mathsf{CP}$) transformation, the partition function is complex conjugated, and then
\begin{align}
    \mathsf{P}:
    Z[A_{1,2}]
    &\mapsto Z[A_{1,2}]^*\notag\\
    &=
    \exp{\bigg(\frac{2\pi\im}{N} \int A_{1} \cup A_{2}
    \bigg)} Z[A_{1,2}].
\end{align}
This is how the anomaly matching of \eqref{eq:CP-ZN_anomaly} is satisfied for the confinement states.

\subsubsection{Oblique confinement phases}
\label{subsubsec:{Oblique confinement phase}}

Oblique confinement phase~\cite{tHooft:1981bkw, Cardy:1981qy,Cardy:1981fd} is characterized by the condensation of the higher magnetically charged dyons, and the $2$d counterpart in our model is the situation where the relevant perturbations are given by the operators with $q>1$, such as $(p,q) = (-1,2)$. 

To be concrete, let us focus on the oblique confinement with $(p,q)=(-1,2)$, and then its scaling dimension is given by 
\begin{align}
    \Delta_{-1,2}
    =
    \frac{N}{2}\bigg[
    \frac{N}{R^2}\left(
    1-\frac{\bar{\theta}}{\pi}\right)^{2}
    + \frac{4R^2}{N}
    \bigg].
\end{align}
When $R^2$ is sufficiently small, the magnetic contribution (i.e. the second term) becomes small. When $\bar{\theta}\approx \pi$, the electric contribution to the scaling dimension is also negligible, and thus the $(p,q)=(-1,2)$ condensate can be relevant in the low-energy limit. 


For this case, the mass gap is generated by $\cos(-N\phi_1+2\tilde{\phi}_2)$ and $\cos(N\phi_2+2\tilde{\phi}_1)$, but 
he symmetry realization of this gapped phase depends crucially on whether $N$ is even or odd. 
For even $N$, the following local vertex operators have nonzero expectation values, 
\begin{align}
    \langle
    \rme^{-\im \left(\frac{N}{2}\phi_{1} -\tilde{\phi}_{2}\right)}
    \rangle
    = \pm v,\quad 
    \langle
    \rme^{\im \left(\frac{N}{2}\phi_{2} + \tilde{\phi}_{1}\right)}
    \rangle
    =
    \pm v,
\end{align}
for some $v>0$, and there are $4$-fold degenerate ground states associated with the spontaneous symmetry breaking, 
\begin{align}
    \mathbb{Z}_{N}\times \mathbb{Z}_{N}
    \xrightarrow{\mathrm{SSB}}
    \mathbb{Z}_{N/2}\times \mathbb{Z}_{N/2}.
\end{align}
We can check that the anomaly matching condition at $\bar{\theta}=\pi$ is satisfied in the oblique confinement phase~\cite{Honda:2020txe}. 
For confirm it, it is sufficient to show that $\mathbb{Z}_{N/2}\times\mathbb{Z}_{N/2}$ and $\mathsf{P}$ does not have an 't Hooft anomaly. 
Let $\tilde{A}_a$ denote the background gauge field for $\mathbb{Z}_{N/2}\times \mathbb{Z}_{N/2}$, then the $\mathbb{Z}_N$ gauge fields are obtained as $A_a=2\tilde{A}_a$ and the anomalous relation~\eqref{eq:CP-ZN_anomaly} becomes 
\begin{align}
    \mathsf{P} :
    Z_{\bar{\theta}=\pi}[2\tilde{A}_{1,2}]
    \mapsto
    \exp{\bigg(
    \frac{2\pi \im}{N} \int (2\tilde{A}_{1})\cup (2\tilde{A}_{2})
    \bigg)}
    Z_{\bar{\theta}=\pi}[2\tilde{A}_{1,2}].
\end{align}
Since $\tilde{A}_a$ are the $\mathbb{Z}_{N/2}$ gauge field, its $2$d topological local counterterm is given by  
\begin{align}
    \frac{2\pi \im k}{N/2}\int \tilde{A}_{1} \cup \tilde{A}_{2},
\end{align}
with some $k = 0,1, \cdots N/2-1 \; \text{mod} \; N/2$, 
and we can always cancel the anomalous phase by setting $k = 1 \; \text{mod}\; N/2$. 

When $N$ is odd, the oblique confinement phase is a trivial phase but nontrivial as an SPT state with the level $\frac{N+1}{2}$ as in the case of the $4$d counterpart~\cite{Honda:2020txe}. 
In fact, the the global symmetry at $\bar{\theta} = \pi$ does not have a genuine 't Hooft anomaly when $N$ is odd, and thus the realization of the trivial gapped phase is consistent with the symmetry constraint. 
The anomalous relation \eqref{eq:CP-ZN_anomaly} is satisfied by the SPT partition function,
\begin{align}
    Z_{\text{oblique}}[A_{1,2}]
    \propto
    \exp{\bigg(-
    \frac{2\pi \im}{N}\frac{N+1}{2}\int A_{1}\cup A_{2}
    \bigg)},
\end{align}
since $\mathsf{P}:Z_{\mathrm{oblique}}\mapsto Z_{\mathrm{oblique}}^*=\rme^{\frac{2\pi\im}{N}\int A_1\cup A_2}Z_{\mathrm{oblique}}$. 
Indeed, we can construct the string-order parameter whose endpoint is given by 
\begin{equation}
    \rme^{\im\left(-\frac{N+1}{2}\phi_1 + (1+\frac{1}{N})\tilde{\phi}_2\right)}, \quad \rme^{\im \left(-\frac{N+1}{2}\phi_2 - (1+\frac{1}{N})\tilde{\phi}_1\right)},
\end{equation}
using that $\frac{N+1}{2}\in \mathbb{Z}$, and we can confirm the above level for the SPT state. 
We note that the SPT level $\frac{N+1}{2}$ for the oblique confinement is different from the ones of the confinement states at $\bar{\theta}=0$ or $\bar{\theta}=2\pi$, corresponding to the $(0,1)$ condensate and $(-1,1)$ condensate. 
Therefore, as long as the $\mathbb{Z}_{N}\times \mathbb{Z}_{N}$ symmetry is respected, there must be a phase transition between the confinement at $\bar{\theta}=0,2\pi$ and oblique confinement phase at $\bar{\theta}=\pi$.

\section{Summary and discussion}
\label{sec:summary}

In this paper, we revisit the $2$d version of the Cardy-Rabinovici (CR) model, which is a toy model of the lattice $U(1)$ gauge theories to study exotic confinement phases, called oblique confinement phases, associated with the finite $\theta$ angles. 
We reformulate the $2$d CR model using the modified Villain lattice formalism and establish the $\theta$ periodicity for the Witten effect and the strong-weak duality in an exact manner at the finite lattice spacings, which were missing in the original version. 
We determine its phase structure using the power of the
duality, symmetry and anomaly, and the perturbative renormalization group.
Let us explain the details of each result briefly.

First, we constructed the $2$d free compact boson theory with the modified Villain lattice formulation, and showed  exact realizations of the Witten effect and the strong-weak duality in this system. 
By using these results, we reformulated the $2$d CR model by introducing the potential term to satisfy the $\theta$ periodicity and electric $\mathbb{Z}_{N} \times \mathbb{Z}_{N}$ symmetry. 
In this model, we can show the mixed 't~Hooft anomaly (or the global inconsistency) between $\mathsf{P}$ and $\mathbb{Z}_{N}\times \mathbb{Z}_{N}$ symmetry at $\bar{\theta} = \pi$.
We also showed that this model possesses the exact duality relation by combining the $\mathbb{Z}_{N}\times \mathbb{Z}_{N}$ gauging and the lattice strong-weak duality. 

We determine if the phase is gapped or gapless by the perturbative renormalization group when the potential is sufficiently small. 
We then classify the gapped phases from the viewpoint of the $\mathbb{Z}_N\times \mathbb{Z}_N$ symmetry in $2$d QFTs. 
We note that the phases at the couplings $-(\tau/N)^{-1}$ and $\tau+1$ can be determined by the phase at $\tau$ by the gauging operation and the stacking SPT states via duality and anomaly. 
All the gapped phases can be mapped to the Higgs phase (i.e. $\mathbb{Z}_N\times \mathbb{Z}_N$ broken state at $\mathrm{Im}(\tau/N)\gg 1$) via the duality operations, and we can systematically determine the gapped quantum phases of this model.  
We understand the counterpart of confinement phases as $\mathbb{Z}_{N}\times \mathbb{Z}_{N}$-symmetric SPT phases, and we confirm their characterization based on the string-order parameters.

The oblique confinement phase, corresponding to the $(p,q) = (-1,2)$ dyon condensation in $4$d, is also investigated, which appears when $\bar{\theta} \simeq \pi$ and the compact boson radius $R^{2}/N$ is sufficiently small. 
When $N$ is even, the symmetry breaking $\mathbb{Z}_{N}\times \mathbb{Z}_{N} \rightarrow \mathbb{Z}_{N/2}\times \mathbb{Z}_{N/2}$ occurs, which satisfies the anomaly matching.
When $N$ is odd, there is no genuine anomaly at $\bar{\theta} = \pi$, so the oblique confinement phase is permitted to become a trivially gapped phase, which is actually the case. 
Its SPT level is identified from the consistency, and we explicitly confirm it from the string-order parameter. 
As a result, the oblique confinement phase around $\bar{\theta}=\pi$ should be discriminated from the confinement states around $\bar{\theta}=0$ and $2\pi$.

Let us comment on several future directions of this study. 
Since we rely on the perturbative renormalization group to determine the phase diagram, our result is limited to the case with $g\ll 1$. 
It would be interesting to explore the dynamics around $g\sim O(1)$, where the electric and magnetic excitations could compete with one another. 
To study the region of $g\sim O(1)$, we need to develop the numerical technique for this model, but the lattice action~\eqref{eq:2dCR_action} has the imaginary part, which causes the sign problem. 
As demonstrated in Refs.~\cite{Gattringer:2018dlw, Sulejmanpasic:2019ytl, Sulejmanpasic:2020lyq}, the worldline representations provide the manifestly sign-problem-free expression with the nonzero $\theta$ angle, so we need to give its explicit formula and confirm the absence of the sign problem in the case of our model. 
We should also note that our expression of the potential $V(\phi_a,\tilde{\phi}_a)$ contains the infinite sum of the electric-magnetic charges $(p,q)$, and we need to use the $(p,q,\tau)$-dependent coupling to make the lattice theory finite with maintaining the exact $\bar{\theta}$ periodicity and strong-weak duality. 
The other approach for studying the region $g\sim O(1)$ would be the application of the tensor-network methods after reformulating the model with the Hamiltonian formalism, and we should point out that the compact bosons in the Hamiltonian formalisms are recent discussed in Refs.~\cite{Fazza:2022fss, Cheng:2022sgb, Seifnashri:2023dpa}.

Another interesting direction would be an extension of the model to have all the possible gapped phases for the $\mathbb{Z}_N\times \mathbb{Z}_N$ symmetry. 
In the case of the $2$d CR model, any gapped phase can be mapped to the Higgs phase by repeatedly applying the strong-weak duality and the $2\pi$ shift of $\bar{\theta}$. 
However, for example, the spontaneous breaking $\mathbb{Z}_4\times \mathbb{Z}_4\xrightarrow{\text{SSB}}\mathbb{Z}_2\times \mathbb{Z}_2$ without the SPT stacking cannot be mapped to the completely symmetry-broken phase by the $\mathbb{Z}_4\times \mathbb{Z}_4$ gauging and the stacking of the $\mathbb{Z}_4\times \mathbb{Z}_4$ SPT states, and thus the CR model does not realize such a phase. 
The same constraint exists also for the $4$d CR model, which means that all the possibilities of the Wilson-'t~Hooft classification are not exhausted by the CR model, and its extension would give us a hint to understand more details about the strong confinement dynamics. 

Last but not least, the extension of our $2$d CR model to the original $4$d problem is also an important future task. 
In the case of $4$d, the quantization of the topological charge has an extra subtlety related to the spin structure, which causes the statistical transmutation~\cite{Goldhaber:1976dp, Jackiw:1976xx} between the monopoles and dyons. The necessity of the spin structure on the lattice for fermionic gapped phases is discussed in Ref.~\cite{Gaiotto:2015zta}, and we expect that it would be useful to consider its application to the $4$d lattice construction of the CR model extending Ref.~\cite{Anosova:2022cjm} to clarify the subtle issue hidden in the original formulation of the CR model.

\acknowledgments
We thank Yui Hayashi and Tin Sulejmanpasic for useful discussion. 
The work of Y.T. is partially supported by Japan Society for the Promotion of Science (JSPS) KAKENHI Grant No. 23K22489 and by Center for Gravitational Physics and Quantum Information (CGPQI) at Yukawa Institute for Theoretical Physics


\appendix
\section{Calculations of the scaling dimension of general vertex operators}\label{appendix_A}

Here, we give a review for the computation of correlation functions for general vertex operators in $2$d compact boson theory with $\theta$ term on the complex plane $\mathbb{C}\simeq \mathbb{R}^2$. 
In particular, we must calculate the correlation function between the original compact boson and the dual compact boson. First, we consider the original action \eqref{eq:2d_CPT+theta_action}, and calculate the classical solutions for $\phi_{a}$. Then it is convenient to introduce the complex coordinates : $z=x_{1}+ix_{2}, \bar{z}=x_{1}-ix_{2}, \p_{z}=\frac{1}{2}(\p_{1}-i\p_{2}), \p_{\bar{z}}=\frac{1}{2}(\p_{1}+i\p_{2})$. So, the classical solutions for $\phi_{a}$ are given by
\begin{align}
    \phi_{a}(z,\bar{z})
    =
    \varphi_{a}^{+}(z) + \varphi_{a}^{-}(\bar{z}),\label{classical-sol_CPT-boson}
\end{align}
where $\varphi_{a}^{+}(z)$ and $\varphi_{a}^{-}(\bar{z})$ are the holomorphic function and the anti-holomorphic function respectively. From the correlation function for $\phi_{a}$, we gain the relations:
\begin{align}
    \langle
    \varphi_{a}^{+}(z)\varphi_{a}^{+}(0)
    \rangle
    =
    -\frac{1}{2R_a^2}\ln{z},\quad
    \langle
    \varphi_{a}^{-}(\bar{z})\varphi_{a}^{-}(\bar{0})
    \rangle
    =
    -\frac{1}{2R_a^2}\ln{\bar{z}},\label{cor-fuc_hol-anti-hol}
\end{align}
where the other correlation functions are zero. Next, we consider the action after introducing the dual compact boson:
\begin{align}
    &S_{\theta}[\p \phi_{1} ,\p \phi_{2},\tilde{\phi}_{1},\tilde{\phi}_{2}]\notag\\
    &=
    \int_{\mathbb{R}^2}\diff^{2}x\left[\sum_{a=1,2}\frac{R_{a}^2}{4\pi}(\p_{\mu}\phi_{a})^{2}
    +
    \im\frac{\theta}{4\pi^{2}}\varepsilon_{\mu\nu}\p_{\mu}\phi_{1}\p_{\nu}\phi_{2}
    +
    \frac{\im}{2\pi}\sum_{a=1,2}\varepsilon_{\mu\nu}\p_{\mu}\phi_{a}\p_{\nu}\tilde{\phi}_{a}\right],
    \label{eq:2d_CPT+theta+dual_action}
\end{align}
where we regard the vector field $\p\phi_{a}$ and the dual compact boson $\tilde{\phi}_{a}$ are the fundamental fields in this stage. Considering the action \eqref{eq:2d_CPT+theta+dual_action}, the equations of motion for the vector fields $\p_{\mu}\phi_{a}$ are given by
\begin{align}
    R_1^2\partial_{\mu}\phi_{1} + \im\frac{\theta}{2\pi}\varepsilon_{\mu\nu}\partial_{\nu}\phi_{2} + \im \varepsilon_{\mu\nu}\partial_{\nu}\tilde{\phi}_{1}
    &=
    0 , \label{EoM_D-CPT1}\\
    R_2^2\partial_{\mu}\phi_{2} - \im \frac{\theta}{2\pi}\varepsilon_{\mu\nu}\partial_{\nu}\phi_{1} + \im \varepsilon_{\mu\nu}\partial_{\nu}\tilde{\phi}_{2}
    &=
    0.\label{EoM_D-CPT2}
\end{align}
By introducing the complex coordinate and substituting classical solutions \eqref{classical-sol_CPT-boson} to the equations of motions \eqref{EoM_D-CPT1},\eqref{EoM_D-CPT2}, the classical solutions for the dual compact bosons are given by 
\begin{align}
    \tilde{\phi}_{1}
    &=
    R_{1}^2\varphi_{1}^{+}(z)-\frac{\theta}{2\pi}\varphi_{2}^{+}(z)
    - R_{1}^2\varphi_{1}^{-}(\bar{z})-\frac{\theta}{2\pi}\varphi_{2}^{-}(\bar{z}) \quad \text{(up to const.)},
    \label{clas-sol_dualCPT1}\\
    \tilde{\phi}_{2}
    &=
    R_{2}^2\varphi_{2}^{+}(z) + \frac{\theta}{2\pi}\varphi_{1}^{+}(z)
    -R_{2}^2\varphi_{2}^{-}(\bar{z}) + \frac{\theta}{2\pi}\varphi_{1}^{-}(\bar{z}) \quad \text{(up to const.)}.
    \label{clas-sol_dualCPT2}
\end{align}
Then, we can compute any correlation functions on the complex plane using the Wick contraction and the relations \eqref{cor-fuc_hol-anti-hol}, \eqref{clas-sol_dualCPT1}, and \eqref{clas-sol_dualCPT2}. 
For example, the two-point function for the general vertex operator  $\rme^{iN_{1}\phi_{1}+iN_{2}\phi_{2}+i\tilde{N}_{1}\tilde{\phi}_{1}+i\tilde{N}_{2}\tilde{\phi}_{2}}$ is given by 
\begin{align}
    &\big\langle \rme^{\im (N_{1}\phi_{1}+N_{2}\phi_{2}+\tilde{N}_{1}\tilde{\phi}_{1}+\tilde{N}_{2}\tilde{\phi}_{2})(z,\bar{z})}
    \rme^{-\im (N_{1}\phi_{1}+N_{2}\phi_{2}+\tilde{N}_{1}\tilde{\phi}_{1}+\tilde{N}_{2}\tilde{\phi}_{2})(0,0)}
    \big\rangle \notag\\
    &=
    \bigg(
    \frac{1}{|z|^{2}}
    \bigg)^
    {
    \frac{1}{2R_{1}^2}\big(N_{1}+\frac{\tilde{N}_{2}\theta}{2\pi}\big)^{2}+\frac{R_1^2}{2}\tilde{N}_{1}^{2}
    + \frac{1}{2R_2^2}\big(N_{2}-\frac{\tilde{N}_{1}\theta}{2\pi}\big)^{2} + \frac{R_2^2}{2}\tilde{N}_{2}^{2}
    }
    \times
    \bigg(
    \frac{\bar{z}}{z}
    \bigg)^
    {N_{1}\tilde{N}_{1} + N_{2}\tilde{N}_{2}}.
\end{align}

\section{A nontrivial \texorpdfstring{$\pi/2$}{pi/2} rotational symmetry }\label{appendix_B}

In this appendix, we discuss the property of the topological term $Q_{\text{top}}$ defined in Eq.~\eqref{eq:lattice_Qtop} (see also Fig.~\ref{fig:2d CPT-MVlattice-topterm}) under the Euclidean $\frac{\pi}{2}$ rotation. 
For this purpose, we formally consider the infinite square lattice, $\mathbb{Z}\times \mathbb{Z}$, and take the rotational center at the origin $x_*=(0,0)$. 
The topological term~\eqref{eq:lattice_Qtop} explicitly violates the usual $\frac{\pi}{2}$ rotational symmetry, but it instead maintains the non-trivial $\frac{\pi}{2}$ rotational symmetry defined by as follows: 
\begin{itemize}
    \item[(1)] $\frac{\pi}{2}$ rotation of the $\phi_{1}$ and $n_{1,\mu}$ fields around the site $x_{*}=\vec{0}$, 
    \begin{align}
        \phi_1(x)\mapsto \phi_1(x'),\quad  
        (n_{1,1}(x),n_{1,2}(x))\mapsto (n_{1,2}(x'),n_{1,-1}(x')), 
        \label{eq:step1_90rot}
    \end{align}
    with $x'=(-x_2,x_1)$. Here, $n_{a,-\mu}(x)=-n_{a,\mu}(x-\hat{\mu})$. 
    \item[(2)] $\frac{\pi}{2}$ rotation for the $\phi_{2}$ and $n_{2,\mu}$ fields around its dual site $\tilde{x}_{*}=\frac{\hat{1}+\hat{2}}{2}=(\frac{1}{2},\frac{1}{2})$, 
    \begin{align}
        \phi_2(x)\mapsto \phi_2(x'+\hat{1}), \quad 
        (n_{2,1}(x),n_{2,2}(x))\mapsto (n_{2,1}(x'+\hat{1}), n_{2,-1}(x'+\hat{1})).
        \label{eq:step2_90rot}
    \end{align}
\end{itemize}

The topological charge density at the site $x$ is given by 
\begin{align}
    &\varepsilon_{\mu\nu}[
    \partial_{\mu}\phi_{1}(x) + 2\pi n_{1,\mu}(x)
    ]
    [
    \partial_{\nu}\phi_{2}(x+\hat{\mu}) + 2\pi n_{2,\nu}(x+\hat{\mu})
    ]\notag\\
    =
    &[
    \phi_{1}(x+\hat{1})-\phi_{1}(x) + 2\pi n_{1,1}(x)
    ]
    [
    \phi_{2}(x+\hat{1}+\hat{2})-\phi_{2}(x+\hat{1}) + 2\pi n_{2,2}(x+\hat{1})
    ]\notag\\
    -&
    [
    \phi_{1}(x+\hat{2})-\phi_{1}(x) + 2\pi n_{1,2}(x)
    ]
    [
    \phi_{2}(x+\hat{1}+\hat{2})-\phi_{2}(x+\hat{2}) + 2\pi n_{2,1}(x+\hat{2})
    ].
\end{align}
Let us apply \eqref{eq:step1_90rot} and \eqref{eq:step2_90rot} to the topological charge density, then we get
\begin{align}
    \longrightarrow
    &[
    \phi_{1}(x'+\hat{2})-\phi_{1}(x') + 2\pi n_{1,2}(x')
    ]
    [
    \phi_{2}(x'+\hat{2})-\phi_{2}(x'+\hat{1}+\hat{2})-2\pi n_{2,1}(x'+\hat{2})
    ]\notag\\
    -
    &[
    \phi_{1}(x'-\hat{1})-\phi_{1}(x')-2\pi n_{1,1}(x'-\hat{1})
    ]
    [
    \phi_{2}(x'+\hat{2})-\phi_{2}(x')+2\pi n_{2,2}(x')
    ].\label{nontrivial-rotation-top}
\end{align}
By shifting the second term as $x' \rightarrow x'+\hat{1}$, we obtain the original topological charge density at $x'$. 


Moreover, we can check the nontrivial $\frac{\pi}{2}$ rotational symmetry in $2$d CR model. This symmetry transformation acts on the dual variables in the same way as the original ones, e.g. $\frac{\pi}{2}$ rotating $\tilde{\phi}_{1}$ around $x_{*}$, and $\frac{\pi}{2}$ rotating $\tilde{\phi}_{2}$ around $\tilde{x}_{*}$:
\begin{align}
    \tilde{\phi}_1(\tilde{x})\mapsto \tilde{\phi}_1(\tilde{x'}-\hat{1}),\quad \tilde{\phi}_2(\tilde{x})\mapsto \tilde{\phi}_2(\tilde{x'}),
\end{align}
where $\tilde{x'}=x'+\frac{\hat{1}+\hat{2}}{2}$. 
This maintains the topological coupling $\tilde{\phi}_a(\tilde{x}) \varepsilon_{\mu\nu} \Delta_\mu n_{a,\nu}(x)$ as $\varepsilon_{\mu\nu} \Delta_\mu n_{1,\nu}(x)\mapsto \varepsilon_{\mu\nu} \Delta_\mu n_{1,\nu}(x'-\hat{1})$ and $\varepsilon_{\mu\nu} \Delta_\mu n_{2,\nu}(x)\mapsto \varepsilon_{\mu\nu} \Delta_\mu n_{2,\nu}(x')$ under the above $\frac{\pi}{2}$ rotation. 
The local potential term $V$ of the CR model in \eqref{matter-action_2dC-R} remains also unchanged after performing the symmetry transformation: Each term of the potential transforms as
\begin{align}
\left\{
    \begin{array}{l}
      \cos{\big(
    Np\phi_{1}(x)+q\tilde{\phi}_{2}(\tilde{x})
    \big)}\mapsto 
     \cos{\big(
    Np\phi_{1}(x')+q\tilde{\phi}_{2}(\tilde{x'})
    \big)} , \\
      \cos{\big(
    Np\phi_{2}(x)-q\tilde{\phi}_{1}(\tilde{x}-\hat{1}-\hat{2})
    \big)}\mapsto
     \cos{\big(
    Np\phi_{2}(x'+\hat{1})-q\tilde{\phi}_{1}(\tilde{x'}-\hat{2})
    \big)}, 
    \end{array}
    \right. \label{nontrivial-rotation-V}
\end{align}
and it goes back to the original form at $x'$ by shifting the second term as $x'\mapsto x'-\hat{1}$


\section{Concrete calculations of the strong-weak duality on the lattice}\label{appendix_C}

In this section, we calculate the strong-weak duality in Sec.~\ref{sec:exact_duality_free_lattice} on the lattice concretely. The basic policy is to introduce the dual integer gauge fields $\tilde{n}_{a}$ by using the Poisson summation formula, 
\begin{align}
    \sum_{n\in \mathbb{Z}} 
    \rme^{-\frac{R^2}{4\pi}[F+2\pi n]^{2} - \frac{\im}{2\pi} \gamma [F+2\pi n]}
    = 
    \frac{1}{R} 
    \sum_{m\in \mathbb{Z}} 
    \rme^{-\frac{1}{4\pi R^2}[\gamma+2\pi m]^{2}+\im m F}.
    \label{eq:PoissonSum}
\end{align}
To apply this formula, we rewrite the lattice action~\eqref{eq:latticeVillainAction_free} as follows:
\begin{align}
    S_{\mathrm{free},\, \theta}
    &=\frac{1}{4\pi}\sum_{ax\mu,by\nu} (\partial_\mu \phi_a(x)+2\pi n_{a\mu}(x))M_{ax\mu,by\nu} (\partial_\nu \phi_b(y)+2\pi n_{b\nu}(y))\notag\\
    &\quad +\frac{\im}{2\pi}\sum_{ax\mu}(\varepsilon_{\mu\nu}\partial_\nu \tilde{\phi}_a(\tilde{x}-\hat{\nu}))(\partial_\mu \phi_a(x)+2\pi n_{a\mu}(x)), 
\end{align}
where 
\begin{equation}
    M_{ax\mu,by\nu}=
    \begin{pmatrix}
        R_1^2 \delta_{\mu\nu} \delta_{xy} & \im\frac{\theta}{2\pi}\varepsilon_{\mu\nu}\delta_{x+\hat{\mu},y}\\
        -\im\frac{\theta}{2\pi}\varepsilon_{\mu\nu}\delta_{x,y+\hat{\nu}} & R_2^2 \delta_{\mu\nu}\delta_{xy}
    \end{pmatrix}. 
\end{equation}
Its inverse matrix is give by replacing the couplings by dual couplings~\eqref{new-Coupling}, 
\begin{equation}
    (M^{-1})_{ax\mu,by\nu}=
    \begin{pmatrix}
        \tilde{R}_1^2 \delta_{\mu\nu} \delta_{xy} & \im\frac{\tilde{\theta}}{2\pi}\varepsilon_{\mu\nu}\delta_{x+\hat{\mu},y}\\
        -\im\frac{\tilde{\theta}}{2\pi}\varepsilon_{\mu\nu}\delta_{x,y+\hat{\nu}} & \tilde{R}_2^2 \delta_{\mu\nu}\delta_{xy}
    \end{pmatrix}. 
\end{equation}
By applying the Poisson summation formula~\eqref{eq:PoissonSum} with introducing the new $\mathbb{Z}$-valued field $m_{a\mu}(x)$, we obtain the dual action as 
\begin{align}
    &\quad \frac{1}{4\pi}\sum_{ax\mu,by\nu}(\varepsilon_{\mu\rho}\partial_\rho \tilde{\phi}_a(\tilde{x}-\hat{\rho})+2\pi m_{a\mu}(x))M^{-1}_{ax\mu,by\nu}(\varepsilon_{\nu\sigma}\partial_\sigma \tilde{\phi}_b(\tilde{y}-\hat{\sigma})+2\pi m_{b\nu}(y))\notag\\
    &-\im \sum_{ax\mu}m_{a\mu}(x) \partial_{\mu}\phi_a(x). 
\end{align}
By rewriting the dual gauge field as $m_{a\mu}(x)=\varepsilon_{\mu\nu}\tilde{n}_{a\nu}(\tilde{x}-\hat{\nu})$, the straightforward computation gives $\tilde{S}_{\mathrm{free}}$ in \eqref{dualization-PF}.

\bibliographystyle{utphys}
\bibliography{./refs}
\end{document}